\documentclass{aa}
\usepackage[utf8]{inputenc}
\usepackage{natbib}
\usepackage[colorlinks]{hyperref}
\usepackage[dvipsnames]{xcolor}
\usepackage{ulem}
\usepackage{booktabs}
\usepackage{dirtytalk}
\usepackage{physics}
\usepackage{comment}

\hypersetup{
    citecolor=[RGB]{44, 117, 255},  
    linkcolor=[RGB]{231, 62, 1},
    urlcolor=[rgb]{0,0,1}
}

\usepackage[]{color}

\definecolor{red}{RGB}{255, 30, 0}

\definecolor{blue}{RGB}{0, 165, 255}

\definecolor{green}{RGB}{106,180,153}

\definecolor{violet}{RGB}{212,115,212}

\newcommand{\Planck}{\textit{Planck} }

\def\SNRthreshold{4.5}

\makeatletter
\renewcommand*\aa@pageof{, page \thepage{} of \pageref*{LastPage}}
\makeatother

\title{Characterising galaxy clusters' completeness function in \Planck with hydrodynamical simulations}

\titlerunning{}
\author{
  S.~Gallo\inst{1}\thanks{E-mail:~\tt{stefano.gallo@universite-paris-saclay.fr}},
  M.~Douspis\inst{1}, E.~Soubri\'e\inst{1,2}, L.~Salvati\inst{1} 
}
\institute{
Université Paris-Saclay, CNRS, Institut d’Astrophysique Spatiale, 91405, Orsay, France
\label{inst1}
\and
Institute of Applied Computing \& Community Code (IAC$^3$),
UIB, Spain\label{inst2}
}
\date{\today}

\abstract{
Galaxy cluster number counts are an important probe to constrain cosmological parameters. One of the main ingredients of the analysis, along with accurate estimates of the clusters' masses, is the selection function, and in particular the completeness, associated to the cluster sample one is considering. Incorrectly characterising this function can lead to biases in the cosmological constraints.
In this work, we want to study the completeness of the \Planck cluster catalog, estimating the clusters' probability of detection in a realistic setting using hydrodynamical simulations. In particular, we probe the case in which the cluster model assumed in the detection method differs from the shape and profiles of true galaxy clusters.
We create around 9000 images of the Sunyaev-Zel'dovich effect from galaxy clusters from the IllustrisTNG simulation, and use a Monte-Carlo injection method to estimate the completeness function. We study the impact of having different cluster pressure profiles, as well as that of complex cluster morphologies on the detection process.
We find that the cluster profile has a significant effect on the completeness, with clusters with steeper profiles producing a higher completeness than ones with flatter profiles. We also show that cluster morphologies have small impact on the completeness, finding that elliptical clusters have slightly lower probability of detection with respect to spherically symmetric ones. Finally, we investigate the impact of a different completeness function on a cosmological analysis with cluster number counts, showing a shift in the constraints on $\Omega_m$ and $\sigma_8$ that lies in the same direction as the one driven by the mass bias.
}


\authorrunning{Gallo et al.}

\begin{document}

\maketitle

\section{Introduction}

Galaxy clusters are the largest gravitationally bound objects in the Universe. They form from the highest peaks in the initial density fluctuations and grow through mergers and the accretion of smaller groups and galaxies, driven by their strong gravitational pull \citep{Kravtsov&Borgani2012-cluster_formation_review}. As a result, galaxy clusters provide valuable information about the growth of cosmic structures and help constrain parameters in the cosmological model \citep{OukbirBlanchard1992, Henry1997, Allen2011}. The abundance of galaxy clusters as function of mass and redshift (cluster number counts), for example, is particularly sensitive to the cosmic matter density and density fluctuations ($\Omega_m$ and $\sigma_8$), as well as the dark energy equation of state. Therefore, galaxy clusters are considered fundamental tools for understanding the universe, and have been used as probes in  numerous studies \citep[e.g.][]{Rozo2010-SDSS_cluster_cosmology, PLANCK2016-PSZ2_cosmology, Pacaud2018-XXLcosmology, Bocquet2019-SPT_SZ_cluster_cosmology, Costanzi2021-DESY1_SPT_cluster_cosmology}.

For these reasons, there have been in the last years considerable efforts to build large galaxy cluster catalogs for cosmological analyses, exploiting their multi-component nature to detect them at different wavelengths. For example, in optical, the Dark Energy Survey \citep{Abbott2020-DES}; in X-rays, the XXL survey \citep{Pierre2016-XMM-XXL, Pacaud2018-XXLcosmology} and eROSITA \citep{Liu2022-eROSITA-clusters}; and at millimeter wavelengths, the Atacama Cosmology Telescope (ACT) \citep{Hilton2021-ACT-SZcatalog}, the South Pole telescope (SPT) \citep{Bleem2015-SPT-SZcatalog} and the \Planck survey \citep{PLANCK2014-PSZ, PLANCK2015-PSZ2}.

In order to be able to extract cosmological information from a cluster survey, it is fundamental to know its \textit{selection function}. The selection function is indeed a key element for any statistical study done with a survey, since it connects the detected objects with the underlying true population in the survey area, characterising the relation between the two sets. It is a function of the cluster properties, and depends on the characteristics of the survey, as well as the detection strategy. It can be divided in two separate functions: the \textit{purity}, which is the probability that a given detection corresponds to a real object, and the \textit{completeness}, the probability that an object in the real population will be detected in the survey.

In particular, the selection function is one of the main ingredients for cosmological analyses with galaxy cluster number counts, since it provides, as function of the cluster observables, an estimate of the fraction of objects detected over the ones actually present in the sky. This is very important information when comparing the number of clusters observed with the one predicted from the theory. 
It is then clear why an accurate characterisation of the selection function is important: an incorrect estimation could lead to biases in the cosmological parameters' constraints. 
Another source of uncertainty when probing cosmology with galaxy clusters is the mass calibration. In fact, cluster masses are not observable, so one has to use other cluster properties that correlate with mass as proxies. The relation between observable and mass is modeled via a statistical scaling relation, which is calibrated using multi-wavelength observations. Nonetheless, uncertainties or mischaracterisation of scaling relation parameters can have an important impact on the results of a cosmological analysis.
Therefore, a better understanding of these sources of bias and their possible correlations is particularly interesting, especially in light of the reported tension in the value of $\sigma_8$ (the amplitude of matter fluctuations) between clusters and CMB estimates \citep{PLANCK2014-PSZ_cosmology, PLANCK2016-PSZ2_cosmology, Bocquet2019-SPT_SZ_cluster_cosmology}.

The selection function is necessarily survey-specific. In this study, we aim to characterise the selection function for cosmological analyses done with galaxy clusters detected via thermal Sunyaev-Zel'dovich (SZ) effect \citep{Sunyaev-Zeldovich1970,Sunyaev-Zeldovich1972,Sunyaev-Zeldovich1980} by the \Planck satellite. For this reason, we take as reference the \Planck \texttt{MMF3} cosmological sample \citep{PLANCK2015-PSZ2}, where \texttt{MMF3} is the detection algorithm used, based on the matched multi-filter technique \citep[MMF,][]{Herranz2002-MMF,Melin2006-MMF}. This sample contains 439 clusters, with masses $\in [0.8, 14.7]\times 10^{14}\, M_\odot$ and redshifts $\in [0.01, 0.97]$, covering $65\%$ of the sky. Due to its very high purity ($>99.8\%$), in the rest of the article we will focus exclusively on the completeness function.

The completeness associated to this survey was already studied by the \Planck Collaboration, and detailed with the various catalogs releases \citep{PLANCK2011-ESZ,PLANCK2014-PSZ,PLANCK2015-PSZ2}. In these works, the completeness is first estimated assuming Gaussian noise of the SZ signal, obtaining a rather simple analytical form for the function. This was then compared with a more direct approach, which relies on mock observations, obtained injecting a population of simulated clusters into the real sky maps. With full knowledge of the \say{true} (injected) cluster population, it becomes possible to compare it to the output of the detection algorithm run on these mock maps. In this way, one aims to reproduce as faithfully as possible the (unknown) conditions of the real detection task, in order to get the best estimate of the completeness, including all the contaminant effects that might be difficult to model analytically.

This approach was carried out in \cite{PLANCK2014-PSZ,PLANCK2015-PSZ2} using simulated SZ signals that assumed spherical symmetry for the clusters, finding substantial agreement with the analytical completeness. The latter was then incorporated as the baseline estimate of the completeness in cosmological analyses of cluster number counts with the \Planck SZ catalog \citep[e.g. ][]{PLANCK2014-PSZ_cosmology,PLANCK2016-PSZ2_cosmology}.

In this paper, we take these works as starting point, and re-approach the analysis to characterise the completeness, focusing on the case in which the assumed cluster model in the detection algorithm is different from the ``true'' injected cluster signals. This situation is to be expected, to an extent, given that a template is by necessity a simplification, constructed to match as well as possible the ``average'' features of a selected cluster sample. 
In particular, the \Planck cluster model is based on the assumption of spherical symmetry, and a single pressure profile is assumed. 

Galaxy clusters, though, are known from both simulations and observations to be generally not spherical \citep{Limousin2013}, due to various dynamical effects such as mergers and asymmetric accretion through cosmic filaments \citep{Gouin2019,Gouin2022, Vallesperez2020}. Departure from spherical symmetry is a first clear difference with the detection template which can possibly bias cluster detection, and therefore the effect of realistic morphology needs to be tested while characterising the completeness. 
A second difference between real clusters and the detection template might come from the pressure profile. First of all, not all clusters have exactly the same profile; there are variations due for example to the cluster dynamical state, which induce a scatter around the average profile of the population. Moreover, the average profile might also differ from the one assumed in the template. This might happen, for example, if the profile assumed in the detection was measured from a biased sample of clusters.

Some tests in this direction were performed in \cite{PLANCK2015-PSZ2}. To probe the effect of cluster morphology, a modest sample of hydrodinamically simulated clusters was used, with fixed angular scale larger than the \Planck beam, where the effect of cluster morphology is supposed to be most relevant. No significant difference was found in the completeness using either realistic or spherical morphologies. Regarding the profile scatter, \cite{PLANCK2015-PSZ2} show that completeness computed from cluster images with pressure profiles scattered around the one assumed in the detection is generally consistent with the analytical completeness estimate, but report a widening effect in the completeness drop-off.

In this work, we study these effects in a comprehensive way, analysing the impact of having a non-perfectly-matching cluster model as template for the matched filter detection technique.

To do so, we use a sample of clusters from a large volume, state-of-the-art hydrodynamical simulation, in a somewhat agnostic way. We extract clusters from the simulation at different redshifts, and produce images of their SZ signal as it would be seen on the sky, based on each cluster's redshift and gas distribution. In this way, when computing the completeness, we automatically include any possible redshift dependence of the cluster properties and dynamical states, as well as the effect due to cluster morphologies, in order to increase the realism of the completeness estimation.

In Section \ref{sec:data} we present the sets of cluster images and the \Planck sky maps used for the completeness analysis. In Sec. \ref{sec:methods}, we describe the MMF detection method and discuss the completeness function and the ways it is estimated in detail. The results on the completeness from our realistic cluster images are presented in Sec. \ref{sec:results}, where we investigate the impact of the different clusters' profiles and asymmetric morphologies. We discuss these results and their impact on cosmological analyses in Sec. \ref{sec:discussion}, together with the limitations of our methods. Finally, in Sec. \ref{sec:conclusion}, we draw the conclusions and sketch the next steps.

\section{Data}\label{sec:data}

First we describe the creation of the mock galaxy cluster images that will be used to estimate the completeness. Starting alternatively from publicly available simulation data and analytical pressure profiles, we compute the Sunyaev-Zel'dovich signal from galaxy clusters as it would be seen by the \Planck satellite. 
Then, the \Planck sky maps are presented, which we clean from real detections to serve as background for the injection of the mock cluster signals.

\subsection{Cluster SZ images}\label{sec:SZimages}

At millimeter wavelengths, such as the ones observed by the \Planck satellite, clusters of galaxies can be observed through the Sunyaev-Zel'dovich (SZ) effect. The SZ effect is a spectral distortion of the cosmic microwave background (CMB), due to the inverse Compton scattering of CMB photons off energetic electrons in the hot ionized gas in galaxy clusters. 

While traveling through a galaxy cluster, CMB photons can be scattered by hot electrons in the intracluster gas and gain some energy from them. This produces a peculiar distortion in the spectrum of CMB radiation, that can be distinguished and detected, and encodes information about the hot gas distribution. In particular, neglecting relativistic corrections, the amplitude of the SZ effect is proportional to the Comptonisation parameter $y$, which in turn is proportional to the integrated electron pressure $P_e$ along the line of sight:
\begin{equation}\label{eq:compton_y}
    y = \frac{\sigma_T}{m_e c^2} \int P_e(l) \dd l 
\end{equation}
where $\sigma_T$ is the Thomson cross-section, $m_e$ the electron mass and $c$ is the speed of light.

Therefore, with the SZ effect it is possible to probe the pressure distribution of the gas in the intracluster medium, and its signal can be used to detect galaxy clusters. 

In the following, we explain the production of a set of images of SZ signal, starting from the gas content in simulated galaxy clusters from the IllustrisTNG simulations.

\subsubsection{Simulation}

\begin{figure}
    \centering
    \includegraphics[width=0.49\textwidth]{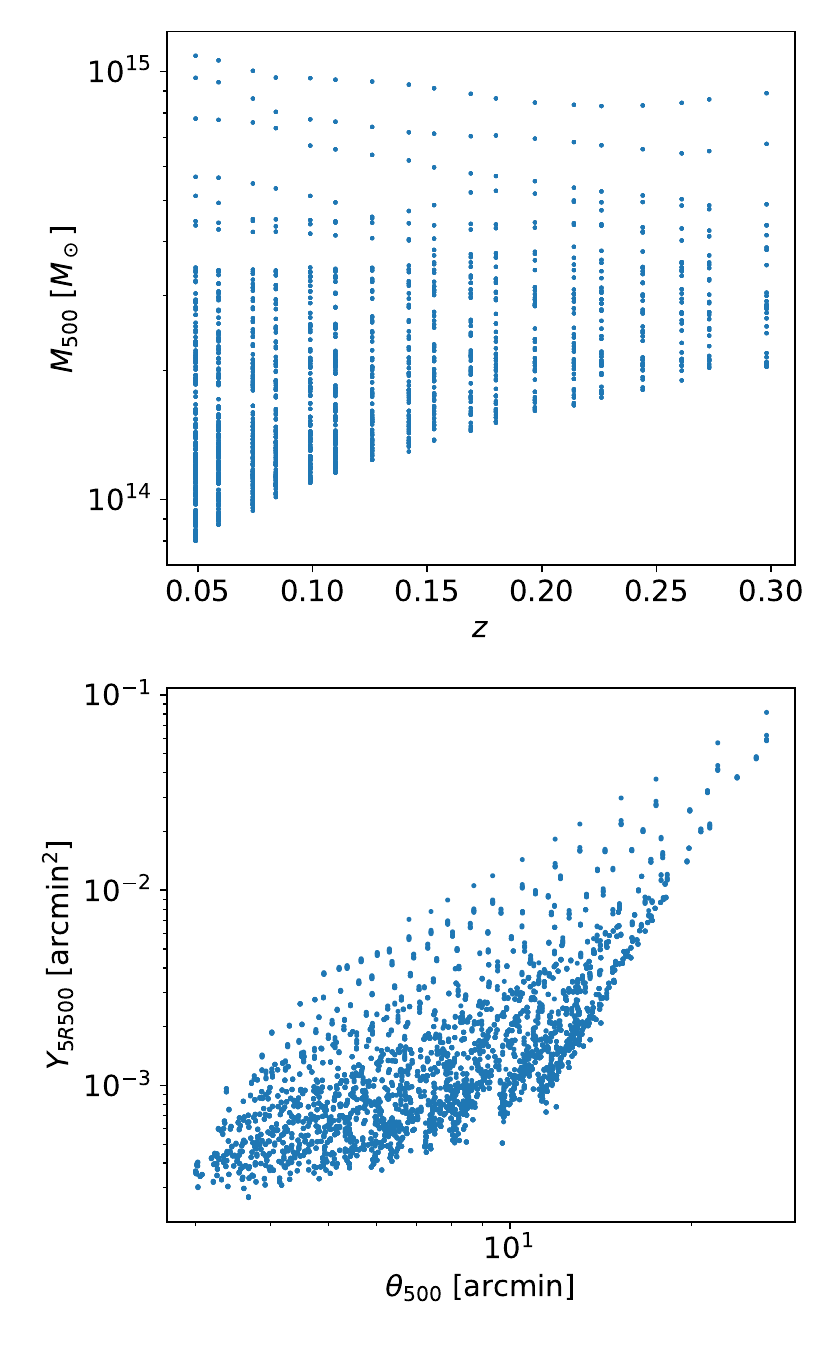}
    \caption{\textit{Top panel}: Mass and redshift distribution of the galaxy clusters selected from the TNG300 simulation, as described in the text. \textit{Bottom panel}: Clusters' distribution in integrated $y$ signal, $Y_{5R500}$, and angular scale, $\theta_{500}$.}
    \label{fig:Mz_Ytheta_distributions}
\end{figure}

IllustrisTNG is a suite of cosmological magneto-hydrodynamical simulations that follow the coupled evolution of dark and baryonic matter through cosmic time, starting from redshift $z=127$ to present time \citep{TNG(a),TNG(b),TNG(c),TNG(d),TNG(e),TNG(f)}. They are run using the moving-mesh code AREPO \citep{springel2010}, and assume cosmological parameters consistent with the results of \cite{PLANCK2016-Cosmology}.
In this work, among the simulations in the suite we focus on the realization called TNG300-1, with a comoving volume of $205\, \mathrm{Mpc}/h$ and a mass resolution for baryons around $7.6 \times 10^6 M_{\odot}/h$. 

The simulation output provides snapshots at different redshifts including a catalog of halos, created running a friends-of-friends (FoF) algorithm \citep{Davis1985} on the dark matter particles, with linking length $b=0.2$. Gas particles are then sorted according to the closest DM particle. 
In this catalog are listed the positions of the halos (identified by the particle with the minimum gravitational potential energy), their masses $M_{500}$ and radii $R_{500}$\footnote{$R_{500}$ is the radius of a sphere centered on the halo within which the average density is 500 times the critical density $\rho_\mathrm{crit}(z)$; $M_{500}$ is the mass contained inside $R_{500}$.}, among other quantities.
Since our goal is to characterise the performance of the MMF detection algorithm, and in particular the completeness, we need a set of clusters that provides a good coverage of the region in mass and redshift where the detection becomes more difficult, and the completeness goes from 1 to 0. From \cite{PLANCK2014-PSZ,PLANCK2015-PSZ2}, we expect this region to be approximately below $\sim 10^{14}\, M_\odot$ at low redshift and $\sim 4\times 10^{14}\, M_\odot$ at high redshift. 
Therefore, we select halos by their mass, with a lower mass limit that depends on redshift, $M_{500}\geq M_\mathrm{min} = \frac{6}{5}\, \qty(4z + \frac{7}{15})\times 10^{14}\, M_\odot$, in the redshift range $0.05 \leq z \leq 0.3$ (which corresponds to 18 snapshots of the simulation). With this selection we obtain a total of 1487 clusters, whose distribution in mass and redshift is shown in Fig. \ref{fig:Mz_Ytheta_distributions}. 

The choice of the redshift range, smaller than the one of the \Planck PSZ2 catalog \citep[$0.01<z<0.97$,][]{PLANCK2015-PSZ2}, is imposed by the limits of the simulation: outside of this range, the cluster distribution in radius and SZ flux doesn't allow to properly sample the completeness function. In particular, at redshift higher than $0.3$ there aren't enough high mass halos, while at low redshift the spacing between the snapshots leaves empty regions in the support of the completeness function. This limit is discussed more in detail in Sec. \ref{sec:discussion}.

Nonetheless, we verified that about $65\%$ of the \Planck PSZ2 cosmological sample falls inside the mass and redshift range covered by our simulation sample.

\subsubsection{Simulation images}

Once the sample of galaxy clusters is selected, we proceed by computing the projected images of Sunyaev-Zel'dovich effect that we will use for the completeness estimation.

First of all, we compute the electron pressure for each of the gas cells associated with the cluster's halos, $P_{e}$. Starting from the cells' densities, electron abundances and internal energies provided in the simulation output, we compute the electron number density $n_{e}$ and temperature $T_e$, and from these the electron pressure as $P_e = k_b n_e T_e$.
Then, we compute images of the Compton-$y$ parameter from the clusters. For each cluster we take six projections: three along the axis of the simulation box, and three along the axis rotated by Euler angles $\qty(\alpha,\beta,\gamma)=\qty(45^{\circ}, 45^{\circ}, 45^{\circ})$. Each image is 4 Mpc wide centered at the cluster position, with a resolution of $256\times 256$ pixels.
In each pixel, the Compton-$y$ parameter is computed as
\begin{equation}
    y = \frac{\sigma_T}{m_e c^2} \int P_e \dd l \approx \frac{\sigma_T}{m_e c^2} \frac{\sum_i P_{e,i} V_i}{A_\mathrm{pix}}
\end{equation}
where $P_{e,i}$ and $V_i$ are the electron pressure and the volume of the $i$th gas cell, respectively, and the index $i$ in the sum runs over all gas cells whose center (provided by the simulation) falls inside the pixel area $A_\mathrm{pix}$.
This rather \say{crude} approximation of the line of sight integral only works when the typical size of the gas cells is smaller than the size of the pixels. In principle, this is true only in the center of the clusters, where density is higher (along with the temperature and therefore pressure). However, in the outer parts of clusters, the cells' sizes are larger but at the same time the pressure is low, making the error we commit by using this approximation negligible. We tested different pixel sizes, and found no appreciable difference in the average cluster $y$ profile (and integrated $y$ signal).

Before being able to inject these images in the Planck frequency maps, they first need to be converted in angular coordinates, as they would be observed on the sky. They are rescaled according to their redshift $z$, using the relations:
\begin{equation}
    \theta_{img} = \frac{4 \mathrm{Mpc}}{d_A(z)}, \qquad
    \theta_{500} = \frac{R_{500}}{d_A(z)}
\end{equation}
where $d_A\qty(z)$ is the angular diameter distance, $\theta_{img}$ is the angular size of the image, and $\theta_{500}$ is the equivalent of $R_{500}$ in angular coordinates, and represents the scale of the cluster. For the rescaled images, we use a pixel size of $0.5\; \mathrm{arcmin}$. This pixel size is therefore lower than the one of the \Planck maps, of about $1.7\; \mathrm{arcmin}$; this is done to avoid having too coarse images before convolving them with the \Planck beams. Later, when the images are injected in the maps, their resolution is adapted to the one of the maps.

To make the images consistent with \Planck observations at the six frequencies of Planck HFI (100, 143, 217, 353, 545, 857 GHz), we convolve them with the corresponding beam, assumed to be circular Gaussian with FWHM taken from \cite{PLANCK2015-PSZ2,Planck2014-HFI_beams}.

Finally, the Compton-$y$ images are transformed in frequency images, multiplying the $y$ parameter by the value of the thermal SZ effect at the 6 frequencies, neglecting relativistic corrections:
\begin{equation}\label{eq:SZ_spectrum}
    \frac{\Delta T}{T_\mathrm{CMB}} = y \cdot g(\nu)
\end{equation}
where $T_\mathrm{CMB}$ is the CMB temperature, and $g(\nu)$ is the spectral signature of the tSZ effect, integrated over \Planck frequency bandpasses, taken from \cite{PLANCK2016-ymap}. 

In this way we obtain a set of 8922 cluster images at the six \Planck frequencies, each associated with the cluster angular scale, $\theta_{500}$, and the integrated SZ signal within a radius of $5\times\theta_{500}$ from the cluster center, $Y_{5R500}$.

\subsubsection{Circular images}\label{sec:SZimages_circ}

\begin{table}
    \centering
    \caption{Generalised NFW pressure profile parameters of the different sets of spherical images. In order: \cite{Arnaud2010} profile (Standard), \cite{PLANCK2013-profile} profile (Planck), \cite{Pointecouteau2021-profile-PACT} profile (PACT), \cite{Tramonte2023-profile-stacked} profile (Tramonte+23), profile obtained changing the $c_{500}$ of \cite{Arnaud2010} (Peaked), and profile obtained fitting the average proifle from the simulation images (SimFit).}
    \label{tab:gNFW_parameters}
    \begingroup
    \renewcommand{\arraystretch}{1.3}
    \begin{tabular}{c|cccc}
	   \toprule
	   Name & $c_{500}$ & $\alpha$ & $\beta$ & $\gamma$ \\
	   \midrule
       Standard & 1.177 & 1.051 & 5.4905 & 0.3081 \\
       Planck & 1.81 & 1.33 & 4.13 & 0.31 \\
       PACT & 1.18 & 1.08 & 4.30 & 0.31 \\
       Tramonte$+$23 & 2.1 & 2.2 & 5.3 & 0.31 \\
       Peaked & 1.5 & 1.051 & 5.4905 & 0.3081 \\
       SimFit & $5.1 \times 10^{-3}$ & 0.71 & 1.33 & 500 \\
	   \bottomrule
    \end{tabular}

    \endgroup
\end{table}

As a complement to the set of images extracted from the simulation, we also generate different sets of spherically symmetric cluster images, in order to test the consistency of the results obtained with the simulation set and explore their implications.

To best compare the completeness obtained with these sets of images with the simulation one we want their distribution in $(Y_{5R500},\theta_{500})$ to be similar. Therefore, we construct a catalog of $(Y_{5R500},\theta_{500})$, taking all the pairs from the simulation set, and applying a random offset to the two values sampled from a Gaussian with a standard deviation of 5\% of each value, in order to sample the same region but not exactly the same values. These new pairs of values are then used to generate the spherical images. 

The images in the different sets are all constructed starting from a spherical pressure distribution, which models the gas in an ideal galaxy cluster. The form of the pressure profile is the generalised Navarro-Frenk-White (gNFW) \citep{NFW1997,Nagai2007-gNFW, Arnaud2010}:

\begin{equation}
    p\qty(x) \propto \frac{1}{\qty(c_{500}x)^\gamma \qty[1 + \qty(c_{500}x)^\alpha]^{\qty(\beta - \gamma) / \alpha}}
\end{equation}

where $x=r/R_{500}$ is the radius in units of $R_{500}$, and $[c_{500}, \alpha, \beta, \gamma]$ are the parameters that determine the shape of the profile. We use six different sets of parameters to build the images, which are detailed in Table \ref{tab:gNFW_parameters}, four come from observational studies \citep{Arnaud2010, PLANCK2013-profile, Pointecouteau2021-profile-PACT, Tramonte2023-profile-stacked}, and two are artificially constructed to approximate the average profile of the simulation images.

To obtain the SZ images for each of the different sets, the pressure profile is first integrated along one direction and transformed in a $y$ map using Eq. \ref{eq:compton_y}. Then this map is rescaled to match the various $(Y_{5R500},\theta_{500})$ of the catalog described before. From this point, the $y$ images here obtained follow the same steps as the simulation images, namely the convolution with the \Planck beams and the transformation in frequency images by the use of Eq. \ref{eq:SZ_spectrum}.

\subsection{Planck Frequency Maps}

To construct the sky maps in which we will inject our cluster images, we start from the six \Planck HFI frequency maps from the second data release \cite{PLANCK2016-frequency_maps}. These maps are given in \texttt{HEALPix} \citep{Gorski2005-HEALPix} pixelisation scheme, with $N_\mathrm{side} = 2048$.

The choice of using real \Planck maps is done, following \cite{PLANCK2014-PSZ,PLANCK2015-PSZ2}, to ensure the most realistic setting possible for the completeness analysis, including all sources of noise and contaminations present during the real detection process. In this light, we choose the maps that were used for the detection of the second \Planck SZ cluster catalog \citep[PSZ2,][]{PLANCK2015-PSZ2}, from where comes the cluster sample whose completeness we aim to characterise.

Given the fact that the MMF algorithm estimates the noise directly from the input maps, the injection of simulated cluster signals in addition to the real ones already present could change the properties of the noise, and therefore impact the detection. For this reason the maps are first subjected to a cleaning process, that aims to remove the SZ signal from the real clusters. The way a given cluster is removed from the maps is the following: starting from the cluster's integrated SZ flux and angular scale, $(Y_{5R500}, \theta_{500})$, a circular image of the cluster SZ emission at the different maps' frequencies is computed, in the same way as in Section \ref{sec:SZimages_circ}, using the profile from \cite{Arnaud2010}. This cluster emission is then subtracted from the maps at the position of the original detected cluster. The differences between the circular template and the real cluster signal leave a residual contribution, but its impact on the noise estimation is certainly smaller than that of the original cluster, and therefore negligible.

The cleaning proceeds in two steps: the first step consists simply in removing all the clusters from the PSZ2 catalogue, obtaining a first \say{cleaned} version of the maps. In the second step, the MMF detection is run on the new maps, to check for additional signals identified as clusters beyond those contained in the PSZ2 catalogue, with a lower limit of $4.25$ in signal-to-noise ratio (S/N). All new detections obtained in this way are then also removed from the frequency maps, which are now in principle free from any relevant SZ source down to S/N $\sim$ 4.5. 
These are the final cleaned maps we use for the completeness analysis. 

Associated to the clean frequency maps we create a mask that covers the emission from the Galaxy and the Magellanic Cloud, as well as an area of 5 times the beam size around point sources \cite[from PCCS2 catalog ][]{PLANCK2016-point_sources_catalog2}, and regions of CO emission \cite{PLANCK2014-CO_emission}. The final unmasked sky fraction is about $78\%$.

\section{Methods}\label{sec:methods}

\subsection{MMF}\label{sec:MMF}

The matched multi-frequency filter (MMF) algorithm \citep{Herranz2002-MMF,Melin2006-MMF} is a commonly used algorithm for the detection of galaxy clusters through their SZ signal. Notably, it has been used to produce cluster catalogs from \Planck data, along with ACT and SPT data \citep{PLANCK2014-PSZ,PLANCK2015-PSZ2,Hilton2021-ACT-SZcatalog,Bleem2015-SPT-SZcatalog}.
This method is designed combining the prior knowledge on the SZ signal from galaxy clusters, namely the spectral signature and spatial characteristics, to produce an optimal filter that returns the maximal signal-to-noise ratio (S/N) in the presence of a galaxy cluster. While the spectral shape of the SZ signal is well known and (in the non-relativistic regime) universal, for the spatial filter one has to do some approximations; the usual choice is to assume spherical symmetry for clusters, and model the radial pressure profile according to the average profile of observed clusters. Of course, it is known that galaxy clusters are in general not spherical, but rather triaxial in shape, and most have even more complex morphologies, due to anisotropic accretion, mergers and astrophysical processes such as shocks and feedback mechanisms. All this variety in morphological features prevents a perfect match between the cluster SZ signal and the spatial template used, in principle impacting the detection performance. This effect is in practice reduced by the smoothing induced by the instrument's beam, which tends to symmetrise the signal, especially for clusters with scales comparable or smaller than the beam size.

More in detail, our implementation of the MMF algorithm is very similar to one used in the construction of \Planck SZ cosmological catalog \citep{PLANCK2014-PSZ, PLANCK2015-PSZ2}, named \texttt{MMF3}. It uses as cluster model a projected gNFW profile with parameters from \cite{Arnaud2010}, $[c_{500},\gamma,\alpha,\beta]=[1.177,0.308,1.051,5.491]$, with two free parameters: the signal amplitude $Y_{5R500}$ and the cluster scale $\theta_{500}$. The spatial filters are therefore constructed from this model, varying the cluster size on a grid of 40 logarithmically-spaced points from $\theta_{500}=1.059$ to 41.195 arcmin.
When run, the MMF algorithm first divides the \Planck full-sky maps in 546 overlapping square patches of $10^\circ$ side. Each patch is then filtered with the templates, resulting in a set of signal-to-noise ratio maps. The peaks in these maps over a S/N threshold of 3 are our candidate cluster detections. The cluster integrated signal $Y_{5R500}$ is given by the filtered map at the peak position, while the cluster size is assumed to be the scale of the filter that maximises the S/N at the cluster location. Finally, the clusters  are merged in a single full-sky catalogue, by merging all detections with lower S/N that fall inside the $\theta_{500}$ of a certain detected cluster.

The final catalogue of detected sources therefore contains the position of the clusters, their S/Ns, and the estimated $\theta_{500}$ and $Y_{5R500}$.
In addition, the algorithm returns an estimate of the noise for each patch and filter size. It is computed directly from the filtered maps, assuming the SZ effect to be small compared to the other astrophysical signals.

\subsection{Completeness}\label{sec:completeness}

The completeness is the probability that a cluster with a certain $(Y_{5R500}, \theta_{500})$ in the true population will be detected, $P\qty(d \mid Y_{5R500}, \theta_{500})$, given the survey and detection method one is considering (in our case, \Planck and MMF, respectively).

A first approximation, as explained in \cite{PLANCK2014-PSZ}, is to assume the noise on the Compton-$y$ parameter as Gaussian. In this case, the completeness can be determined analytically, and takes the form of an error function
\begin{multline}\label{eq:erf_completeness}
    P\left(d \mid Y_{5R500}, \sigma_{Yi}\left(\theta_{500}\right),  q\right)= \\
     \frac{1}{2}\left[1+\operatorname{erf}\left(\frac{Y_{5R500}-q\, \sigma_{Yi}\left(\theta_{500}\right)}{\sqrt{2} \, \sigma_{Yi}\left(\theta_{500}\right)}\right)\right]
\end{multline}
that depends on the integrated SZ signal, $Y_{5R500}$, the noise in the filtered maps at the scale $\theta_{500}$ in a given patch $i$, $\sigma_{Yi}\left(\theta_{500}\right)$, and the threshold in signal-to-noise ratio, $q$, over which a detection is accepted, here we use $q=\SNRthreshold$. The noise per patch per filter used for the ERF completeness is estimated here from the cleaned \Planck maps, in order to avoid any spurious contribution from the injected cluster signals \citep[in the spirit of][]{Zubeldia2022a-iMMF}.
Equation \ref{eq:erf_completeness} is valid for one patch, so to obtain the completeness for the full sky we average it over all patches, weighted by the area of the sky in each patch that is not covered by the mask.\\

\begin{figure*}[th]
    \centering
    \includegraphics[width=.9\textwidth]{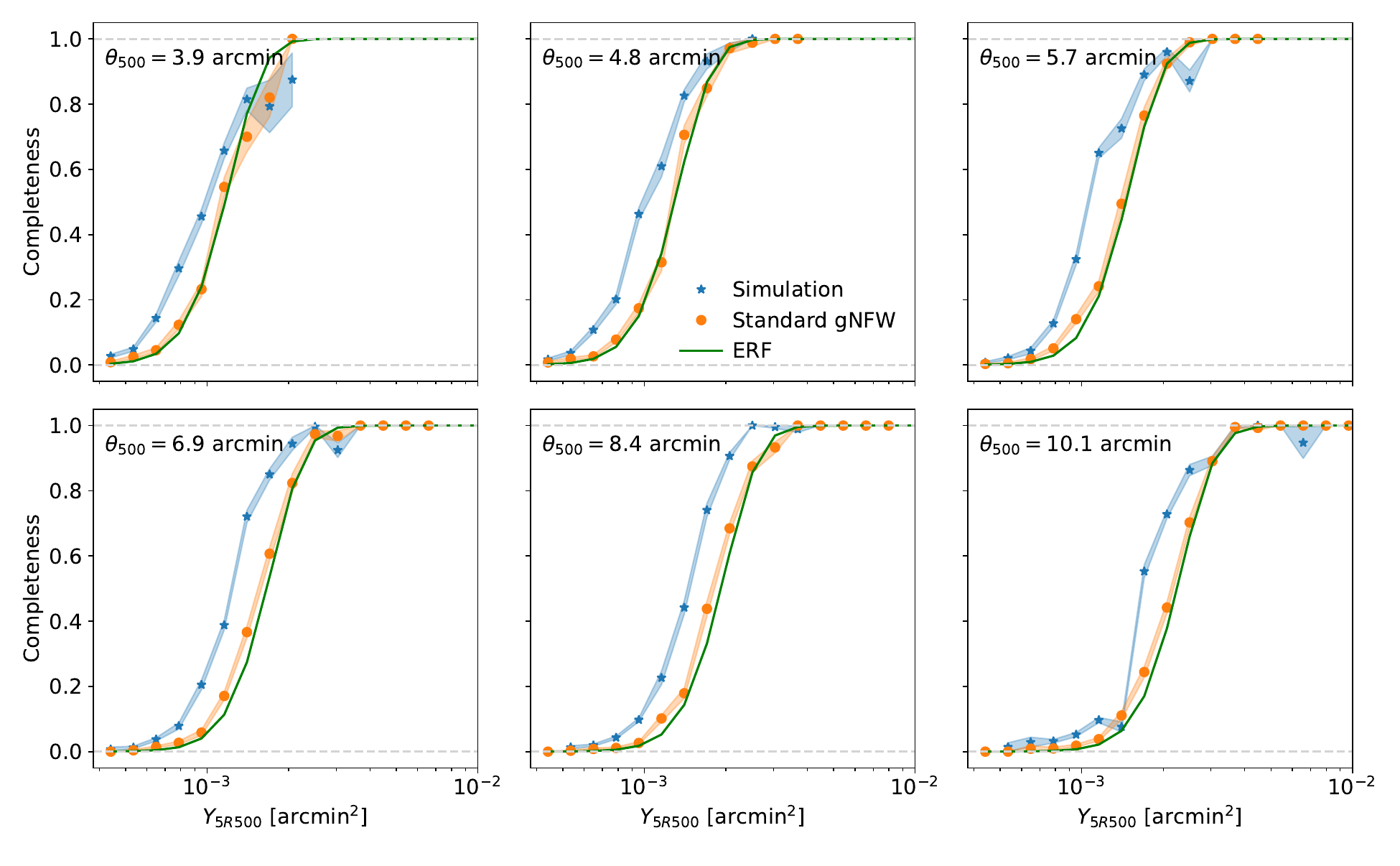}
    \caption{Completeness as function of $Y_{5R500}$, in six $\theta_{500}$ bins, computed with Monte-Carlo injection for the simulation and circular sets of clusters (in blue and orange, respectively), and estimated with the ERF approximation, in green.
    The shaded areas represents the uncertainty on the completeness in each bin, computed via bootstrap resampling.}
    \label{fig:completeness_sims}
\end{figure*}

A more direct way to estimate the completeness function is via Monte-Carlo injection of simulated clusters directly in the sky maps \citep{Melin2005-SZ_selection_function,PLANCK2014-PSZ, PLANCK2015-PSZ2}. 
By injecting the simulated cluster images described in Section \ref{sec:SZimages} in the \Planck maps, and then running the MMF on these new maps, one is able to directly compare the \say{true} cluster set with the catalog of detections returned by the algorithm. This method allows one to estimate the completeness taking into account all the possible sources of biases present in the detection (e.g. algorithmic effects, non-gaussian noise, cluster morphology, etc.), which can be difficult to model analytically.

In order to obtain a sufficient number of detections to properly sample the completeness in the $(Y_{5R500}, \theta_{500})$ plane, without altering the noise properties of the \Planck maps, we create several realizations of injected sky maps, each containing 2000 cluster images. In this way, the average noise of the injected maps differs from the cleaned maps one by less than 5\% (in the scales of interest, below 10 arcmin).

We create 50 mock sky maps for the simulation images set and for the circular images with the Standard profile \citep{Arnaud2010}, while we limit to 10 maps for each of the other sets with different gNFW profiles. 
For each mock map, we select randomly 2000 cluster images, with their associated $Y_{5R500}$ and $\theta_{500}$. The images are then injected in the cleaned \Planck maps, in randomly chosen positions, uniformly distributed outside the Galactic and point sources mask, avoiding overlaps with other injected clusters.
This results in having on average about 4 clusters per ($10^\circ\times10^\circ$) patch. 
The MMF detection algorithm is then run on the maps, obtaining for each a catalog of candidate detections, with candidates' positions, S/Ns, $\theta_{500}$ and $Y_{5R500}$. A threshold in S/N $>\SNRthreshold$ is imposed on the catalogs, the same as for the full \Planck PSZ2 catalog. We choose this threshold for all the completeness tests due to the larger number of detected clusters, which allows for better statistics for the completeness. We checked that the results we obtain are consistent to what we would get choosing S/N $>6$ (as for the \Planck cosmology sample).

Comparing these catalogs with the respective catalogs of injected sources, we determine which clusters among the injected ones are correctly detected by the algorithm.
The matching between the input and output catalogues is done in the following way: a cluster is considered detected if there is an entry in the MMF output catalogue within a radius of 5 arcmin of the injected cluster's position. If there are multiple detections in that area, the one closest to the true position is chosen. Otherwise the cluster is marked as undetected. This procedure is consistent with the one of \cite{PLANCK2014-PSZ,PLANCK2015-PSZ2}.
Then, the resulting single-sky matched catalogues, together with their entries' detection state, are stacked into a single one, which is used to compute the completeness. This step is necessary to obtain good statistics in all $(Y_{5R500}, \theta_{500})$ bins.

Finally, the completeness for each cluster set is computed as the ratio between the number of detected over injected clusters in logarithmically spaced bins of $(Y_{5R500}, \theta_{500})$. If in a bin there are less than 10 clusters, we deem it unreliable and discard it. The errorbars on the completeness are computed via bootstrap resampling of the matched detections table\footnote{This approach gives a useful estimation of the statistical error on the Monte-Carlo mean completeness, but probably underestimates other sources of uncertainty, like the sky-by-sky variations.}.

\section{Results}\label{sec:results}

Here, we present the completeness function estimated with the Monte-Carlo injection method using the set of cluster images from the IllustrisTNG simulation, and discuss its departure from the analytical ERF completeness with the help of the other sets of circular cluster images with varying profiles, described in Section \ref{sec:SZimages_circ}.

\subsection{Completeness with the simulation images}

The completeness function obtained from the simulation images set is shown in Fig. \ref{fig:completeness_sims}, as function of $Y_{5R500}$, in six $\theta_{500}$ bins. It is compared with the analytical ERF completeness, and with the one obtained with the ``Standard gNFW'' images set, that uses the \cite{Arnaud2010} profile (the same profile used to build the filters' template). This set of circular images is meant to act as benchmark for the injection method, and also as test for the ideal case in which the cluster images match the detection template almost perfectly. Thus, a priori, the matched filter should yield the maximum response for these images; and, as consequence, the maximum completeness. We therefore expect the simulation images, that on top of having a variety of different profiles also show complex morphologies that depart from spherical symmetry, to be more difficult to detect; hence producing a lower completeness.

We see that this prediction is not fulfilled by the results of Fig. \ref{fig:completeness_sims}. While the completeness of the ``standard'' images tends to agree well with the analytical ERF \citep[which is in agreement with the results of][]{PLANCK2014-PSZ}, the simulation images produce a completeness that is almost always higher than both other estimates, with differences as high as $0.4$ in some bins. This means that the simulation cluster images, despite their imperfect match with the detection template, have a higher detection probability than the ones for which the match is near perfect.

To help us understand this result, we can imagine to split the contributions of the simulation images to the completeness in two parts: the average $y$ profile, and the non-spherical morphology of simulated clusters. In the following, we study the impact on the completeness of these two aspects separately.

\subsection{Impact of cluster profile}\label{sec:completeness_diffprofiles}

\begin{figure}
    \centering
    \includegraphics[width=.49\textwidth]{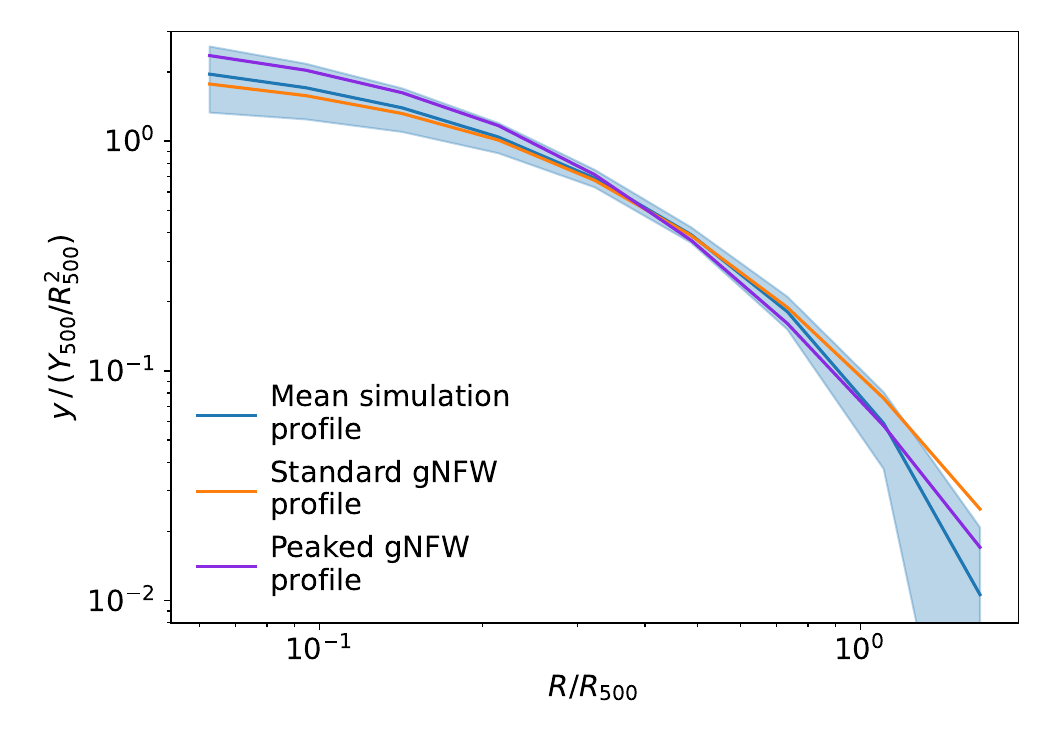}
    \caption{Average Compton-$y$ profile of simulation images  (in blue, shaded area corresponds to standard deviation of profile sample), compared with the $y$ profile obtained from the \cite{Arnaud2010} pressure profile (in orange) and the ``peaked'' profile described in the text (in purple), in units of $R_{500}$ and $Y_{500}/R_{500}^2$. 
    }
    \label{fig:yprofiles}
\end{figure}

\begin{figure}
    \centering
    \includegraphics[width=.49\textwidth]{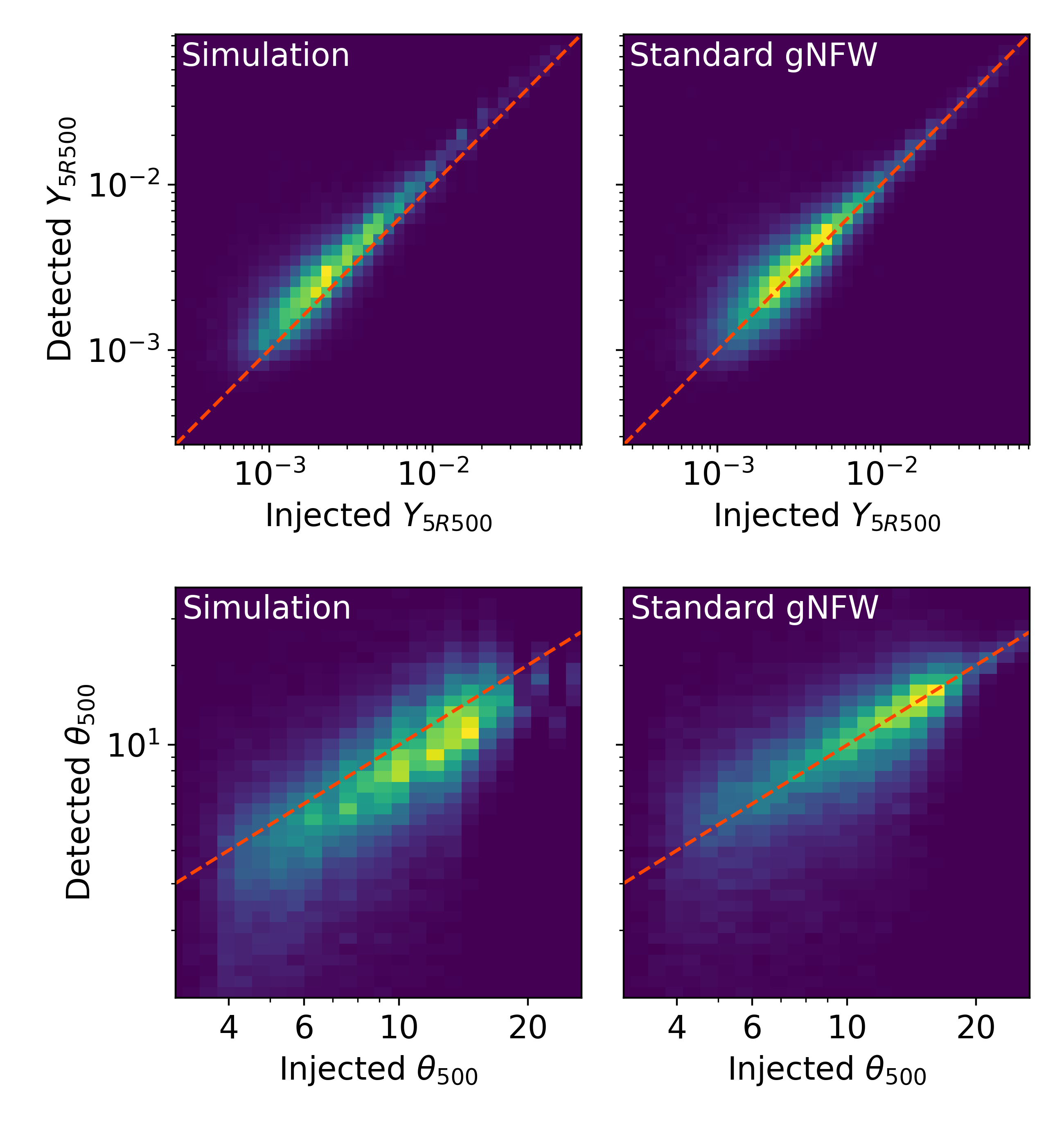}
    \caption{Comparison of real cluster properties vs detected ones. In the top row, the total integrated SZ flux, $Y_{5R500}$; in the bottom row, the cluster radius, $\theta_{500}$. The columns show two different sets of cluster images: simulation images and circular images with \cite{Arnaud2010} profile. We can see that the circular images' properties are correctly recovered on average, but the detection algorithm gives biased estimates of the simulation images' properties: it overestimates $Y_{5R500}$, and underestimates $\theta_{500}$.
    }
    \label{fig:thetaY_recovery}
\end{figure}

In Fig. \ref{fig:yprofiles} we compare the average $y$ profile of the simulation images with the integrated \cite{Arnaud2010} profile. 
We can see how the average simulation profile tends to be overall steeper than the one from \cite{Arnaud2010}, especially in the outer part, from $\sim 0.7 R_{500}$, thus being generally more concentrated, or peaked, with respect to the assumed profile. This trend could be the cause of the higher completeness of the simulation images compared to the Standard gNFW ones, since the MMF algorithm, trying to find the optimal parameter to fit its flatter profile template to the cluster signal, might favour a smaller radius compared to the real one. This underestimation of the radius leads to an increase in S/N, given that the MMF noise estimate increases with filter radius.
This picture is confirmed in Fig. \ref{fig:thetaY_recovery}, where the bias in the detected cluster radii is evident, with a clear underestimation, and a median ratio between detected and injected $\theta_{500}$ of the order of $\sim 1.2$. This bias mostly disappears when analysing the detected $\theta_{500}$ of the Standard gNFW set (with only about $1\%$ median difference), whose profile is the same as that of the MMF template. This indicates that the performance of the detection process depends on the assumed cluster profile in the template. 
Looking at the distribution of the detected SZ signal $Y_{5R500}$ of the two sets of cluster images (Fig. \ref{fig:thetaY_recovery}, top row), we find that both tend to be overestimated with respect to the injected quantities, but with less difference between the two sets. We find an overestimation of about $9\%$ for the Standard gNFW set, consistent with the result of \cite{PLANCK2014-PSZ}, while the Simulation set has a median overestimation of $\sim 25\%$, which leads to an increased signal-to-noise ratio, that in turn contributes to the increase of the completeness.

\begin{figure}
    \centering
    \includegraphics[width=.45\textwidth]{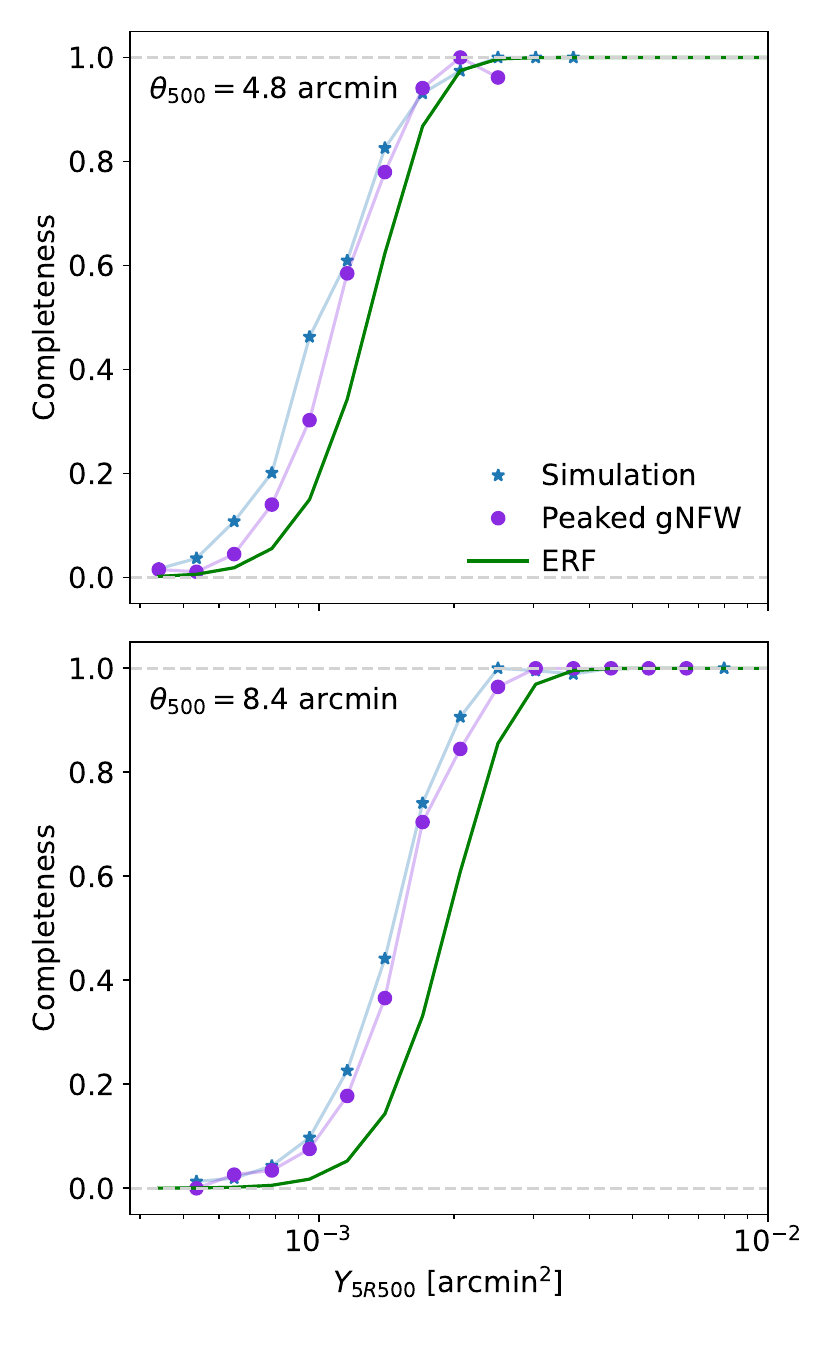}
    \caption{Same as in Fig. \ref{fig:completeness_sims}, comparing the completeness obtained from the simulation images (in blue) with the one from the Peaked profile images (in purple) and the ERF completeness (green).}
    \label{fig:completeness_peaked}
\end{figure}

To test the impact of a different mean profile on the completeness, we use the Peaked gNFW set described in Sec. \ref{sec:SZimages_circ}, which is built from a pressure profile that maintains the same parameters as the \cite{Arnaud2010} one, except for the concentration parameter, $c_{500}=1.5$. 
This higher parameter determines a $y$ profile that is higher in the center and lower around $R_{500}$ than the \cite{Arnaud2010} profile, and roughly reproduces the average profile of the simulated clusters at large radii, as it can be seen in Fig. \ref{fig:yprofiles}. Using this set of images, we compute again the completeness via Monte-Carlo injections and show the results in Fig. \ref{fig:completeness_peaked}, for two $\theta_{500}$ bins as an example. It is apparent how the completeness computed using the images with the Peaked profile agrees better with the one obtained with the simulation images, in particular at larger cluster scales, while at low $\theta_{500}$ it looks more in between the simulation and the ERF completeness. It is a confirmation that a change in the shape of the profile can lead to an increased detection probability, despite the non-perfect match with the filter template.

\begin{figure}
    \centering
    \includegraphics[width=.45\textwidth]{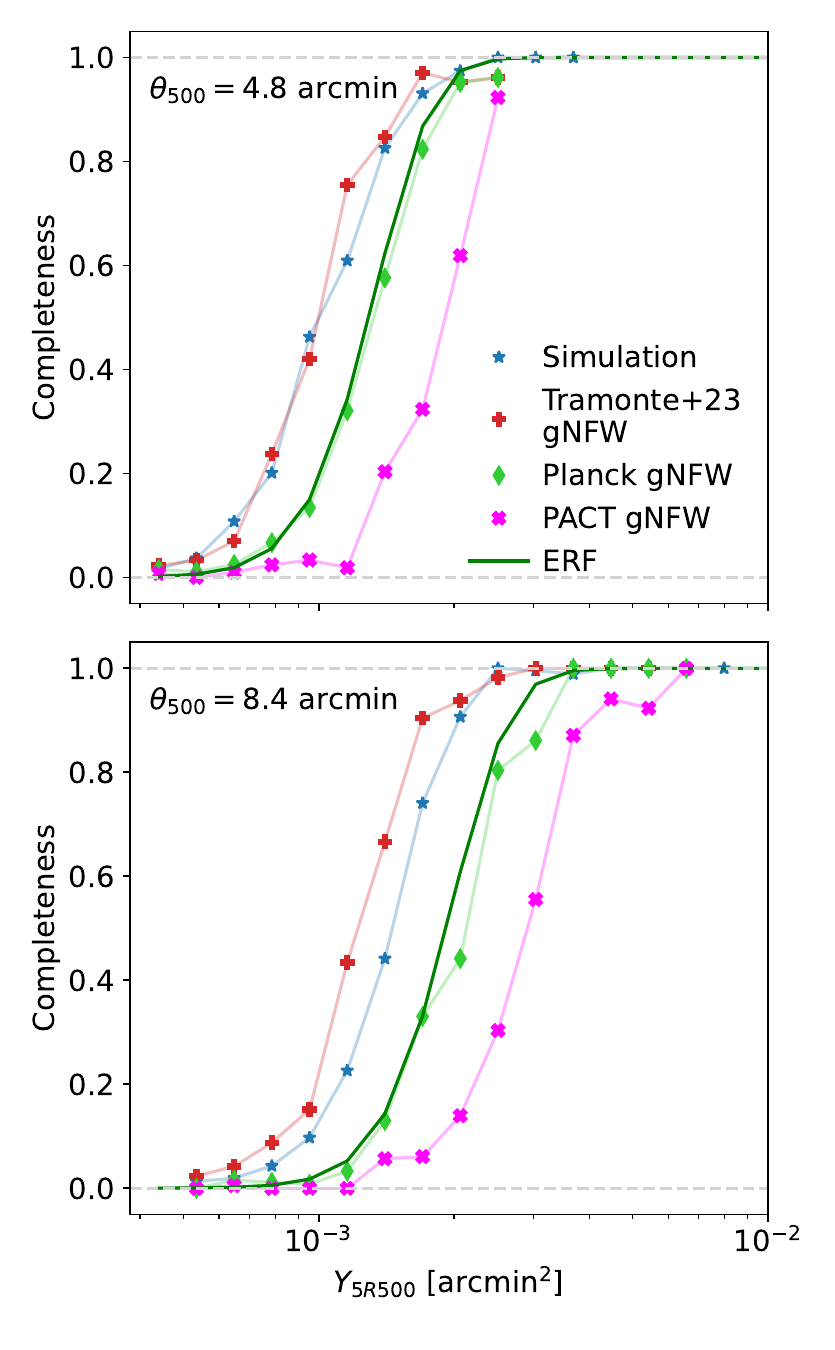}
    \caption{Same as in Fig. \ref{fig:completeness_sims}, comparing the completeness from images with three different observed profiles. \textit{Green}: \cite{PLANCK2013-profile} profile; \textit{pink}: \cite{Pointecouteau2021-profile-PACT} profile; \textit{red}: \cite{Tramonte2023-profile-stacked} profile.}
    \label{fig:completeness_different_profiles}
\end{figure}

\begin{figure}
    \centering
    \includegraphics[width=.49\textwidth]{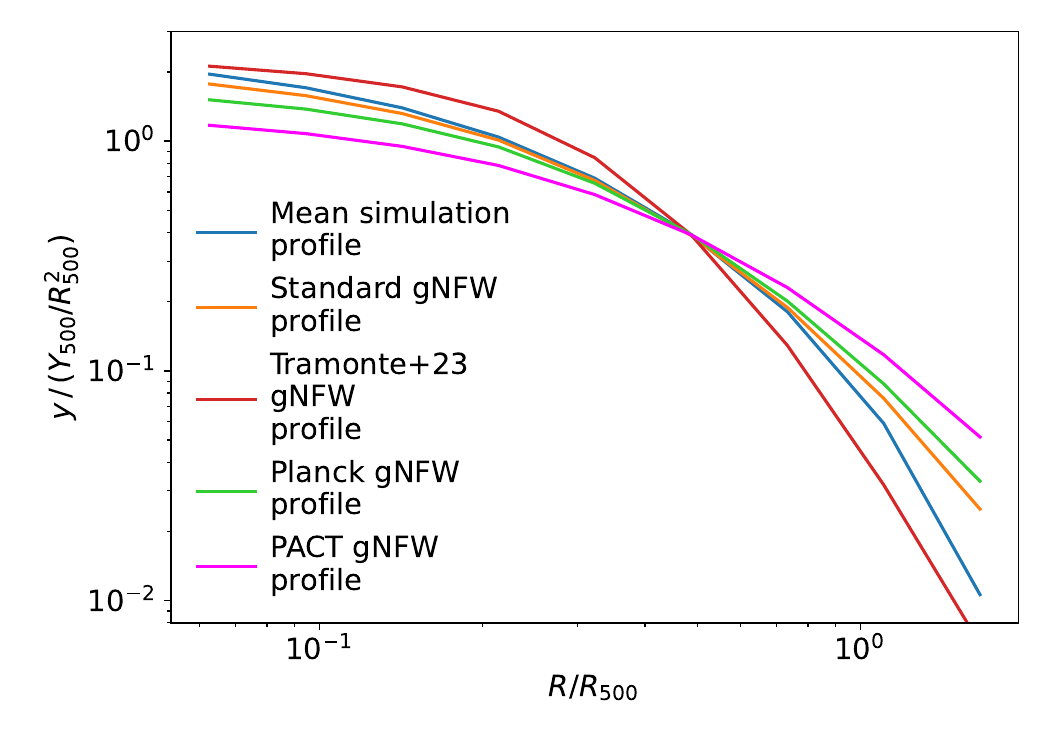}
    \caption{Compton-$y$ profiles in units of $Y_{500}/R_{500}^2$, as function of normalized radius, $R/R_{500}$. \textit{In blue}: mean $y$ profile of the simulation images; \textit{in orange}: profile from \cite{Arnaud2010}; \textit{in green}: profile from \cite{PLANCK2013-profile}; \textit{in pink}: profile from \cite{Pointecouteau2021-profile-PACT}; \textit{in red}: profile from \cite{Tramonte2023-profile-stacked}. }
    \label{fig:y_profiles_sim_gnfw_diffprofiles}
\end{figure}

To probe further this result, we test the impact of using spherically symmetric images with three different gNFW profiles obtained by different studies, fitting the gNFW profile to real observed cluster samples.

The chosen works are: \cite{PLANCK2013-profile}, where the profile was fitted on 62 SZ-selected clusters using Compton-$y$ data from \Planck and x-ray observations from \textit{XMM-Newton}; in \cite{Pointecouteau2021-profile-PACT}, the SZ signal from the combined map of \Planck and ACT \citep{Aghanim2019-PACT} of 31 clusters  was used to constrain the gNFW parameters; finally, \cite{Tramonte2023-profile-stacked}, fitted the $y$ profile obtained by stacking \Planck SZ signal of large number of clusters from different surveys, in different redshift bins. From this study we take the parameters obtained from 4421 clusters in the $z<0.35$ redshift bin, which overlaps with our simulation sample. 
Of these profiles, shown in Fig. \ref{fig:y_profiles_sim_gnfw_diffprofiles}, the first two are flatter than the \cite{Arnaud2010} one (slightly the first, more the second), while the third is flatter in the inner cluster region (up to $\sim 0.1 \, R_{500}$) and then gets steeper than all the other profiles. With this choice, we test the impact of the scatter in the observed profiles in both the flatter and steeper directions.

The results of this test are shown in Fig. \ref{fig:completeness_different_profiles}, for two cluster scales. The effect of using different profile parameters is consistent with our previous results: a steeper profile tends to increase the completeness at all scales, while a flatter one generally decreases it. More in detail, one can notice how the profile of \cite{PLANCK2013-profile}, that is the closest to the detection template, gives a completeness that is comparable with the theoretical estimate, although it tends to be slightly lower for the larger cluster scale. 
One a similar note the \cite{Pointecouteau2021-profile-PACT} profile, the flattest of the three we use, has remarkably lower completeness than the ERF curve. On the other hand, the profile from \cite{Tramonte2023-profile-stacked} gives a completeness that is generally higher than the ERF, like for the simulation set. This further confirms the hypothesis of the relation between the shape of the pressure profile and the estimated completeness, in particular pointing towards a greater importance of the slope outside the very center of the cluster, although this fact needs to be further evaluated.

\subsection{Impact of cluster asymmetry}

\begin{figure*}
    \centering
    \includegraphics[width=.9\textwidth]{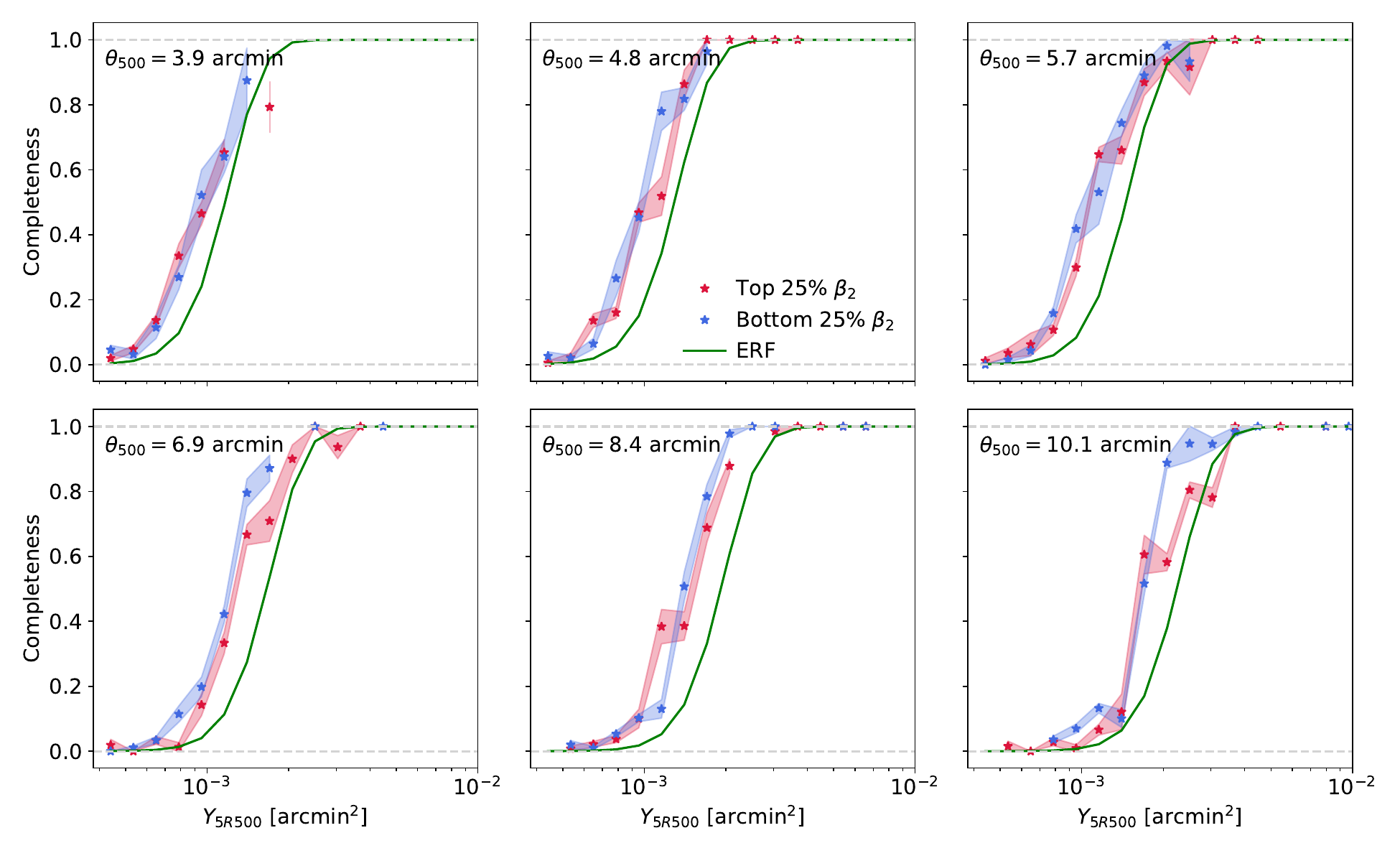}
    \caption{Same as in Fig. \ref{fig:completeness_sims}, comparing the completeness of two subsets of the the simulation images. In red images with the 25\% highest $\beta_2$ (more elliptical), in blue images with the 25\% lowest $\beta_2$ (more spherical).}
    \label{fig:completeness_beta2}
\end{figure*}

\begin{figure}
    \centering
    \includegraphics[width=.49\textwidth]{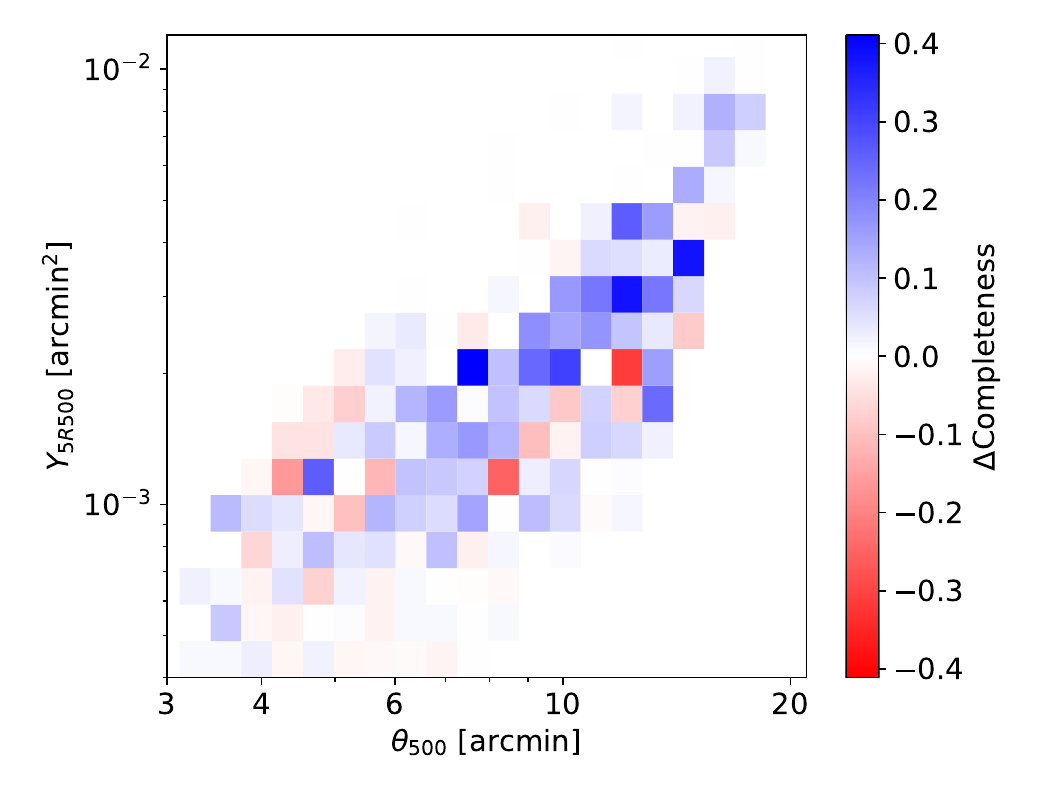}
    \caption{Difference of completeness between the ``more spherical'' and ``more elliptical'' subsets of the simulation images, in bins of ($\theta_{500}$, $Y_{5R500}$). The bins in blue are the ones in which the ``more spherical'' completeness is higher, in red when is the ``more elliptical'' that dominates.}
    \label{fig:diff_completeness_beta2}
\end{figure}

\begin{figure}
    \centering
    \includegraphics[width=.45\textwidth]{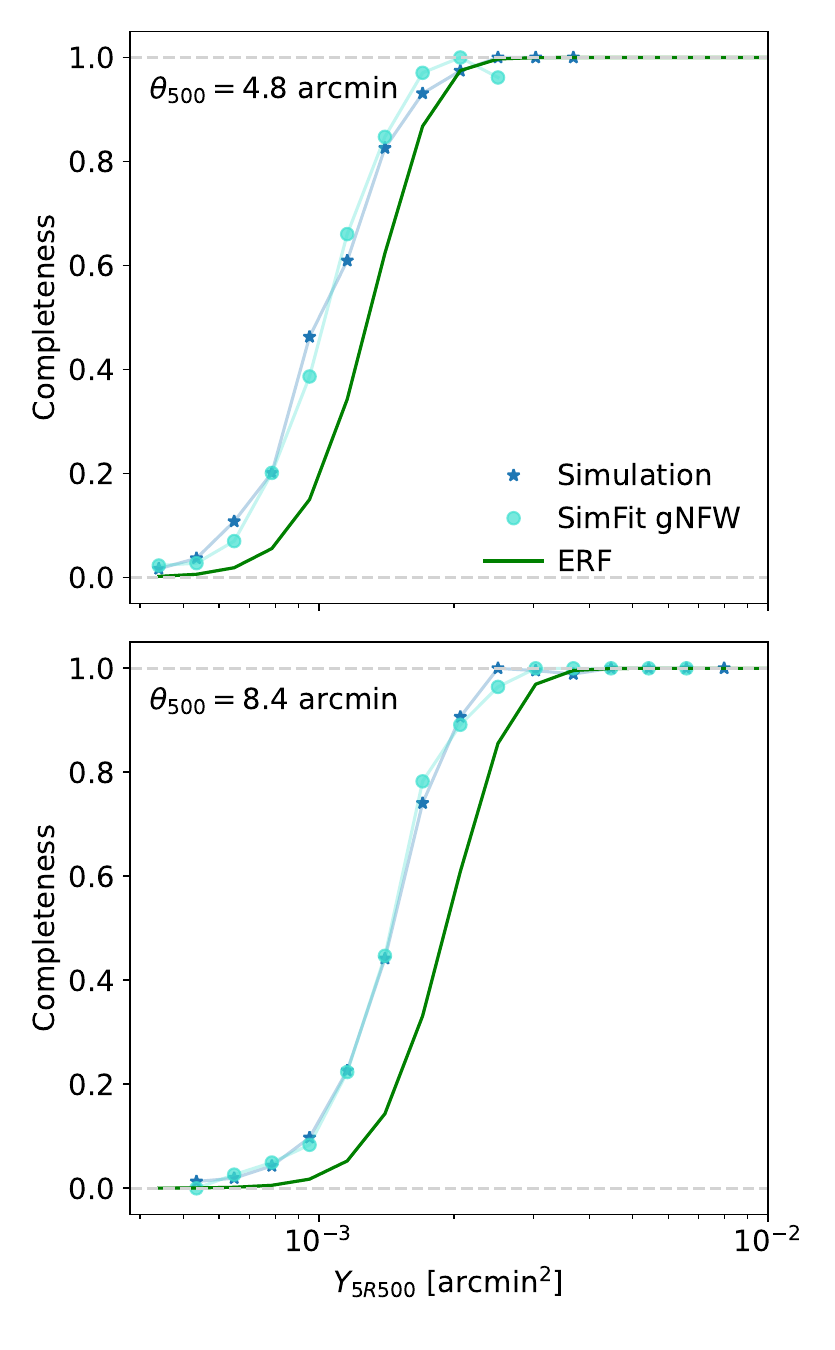}
    \caption{Same as in Fig. \ref{fig:completeness_sims}, comparing the completeness obtained from the simulation images (in blue) with the one from spherical images with profile fitted to the average simulation profile (light blue).}
    \label{fig:completeness_simFit}
\end{figure}

We also study the impact of the departure from spherical symmetry of the cluster images on the completeness. 
Following \cite{Gouin2022}, we quantify the degree of asphericity using a method based on 2D multipole moments decomposition \citep[first introduced by][]{schneider1997}. The method consists in a decomposition of the projected cluster gas distribution $\Sigma(R, \phi)$ in harmonic modes $m$, integrated in a radial aperture:
\begin{equation}
     Q_m (\Delta R) = \int^{R_{max}}_{R_{min}} \int_0^{2\pi}   R \ dR  \ d \phi \ e^{im\phi} \ \Sigma (R,\phi) \, .
     \label{eq:multipole_qm}
\end{equation}
Each mode highlights different patterns in the gas distribution, with higher $m$ corresponding to smaller angular scales. Then, to quantify the relevance of each mode with respect to the spherical symmetry, the multipolar ratio $\beta_m$ is defined \citep{Gouin2022} as
\begin{equation}
    \beta_m =\frac{ \vert Q_m \vert }{ \vert Q_0 \vert} \,.
    \label{eq:multipole_beta}
\end{equation}
Here, we apply this method on the Compton-$y$ images from the simulation (before the convolution with the \Planck beams), in the radial aperture inside $\theta_{500}$.
In particular, we concentrate on the $\beta_2$ multipolar ratio, which has been shown to correlate with the ellipticity of the cluster gas and is expected to be the leading non-spherical mode inside clusters \citep{Gouin2022}, so we decide to use it as proxy for the clusters' asphericity. 

Once we obtain the value of $\beta_2$ for each simulation image, two groups of images are selected, the \say{more elliptical} and \say{more spherical} subsets, formed by the images with the 25\% highest and lowest values of $\beta_2$ respectively, for a total of 2231 images per group.
The completeness for the two subsets are then computed, and shown in Fig. \ref{fig:completeness_beta2}, with the analytical ERF completeness as reference. 

Comparing the completeness of the more and less spherical images, one can see no appreciable difference between the two curves for cluster sizes below $\sim 6$ arcmin. This is expected, since the \Planck beam size is indeed about 6 arcmin on average (and goes up to 10 arcmin for the 100 GHz channel), effectively erasing most smaller-scale morphological differences and symmetrising the images. Above 6 arcmin, instead, we start seeing a small difference between the two subsets, with the more elliptical images having on average lower completeness than the more spherical ones, and the difference appears to become more important with larger cluster radii. This trend can be more clearly visualised in Fig. \ref{fig:diff_completeness_beta2}, where the difference between \say{more elliptical} and \say{more spherical} images' completeness is shown, in bins of $\theta_{500}$ and $Y_{5R500}$. Indeed, the completeness difference tends to increase with increasing $\theta_{500}$, a sign that for resolved clusters the morphology has an impact on the detection probability (i.e. the completeness), although this impact remains moderate, at least in the range of scales probed by our cluster sample.

A further test on the impact of realistic cluster morphologies in the completeness estimation was performed, comparing the simulation set with a set of circular images with the same profile as the average one from the simulation (named SimFit profile in Table \ref{tab:gNFW_parameters}). 
The two completeness functions are shown in Fig. \ref{fig:completeness_simFit}. 
We see that the two curves overlap almost perfectly, suggesting that the $y$ profile is the dominant drive for the completeness, while the small effect of the cluster morphology remarked in the previous paragraph is not visible anymore, probably washed out by considering the full cluster population. So, we are led to believe that, in the context of \Planck and in the range of scales probed by our simulation cluster set, the clusters' departure from spherical symmetry does not induce an appreciable bias in the completeness function. Nevertheless, we show that there is an effect related to cluster morphologies, when separating the ``most elliptical'' images from the ``most spherical'' ones. This effect could be larger in a context where, for example, the beam size is smaller, which will be interesting to check in a future work.

\subsection{Impact on Cluster Counts Cosmology}

\begin{figure}
    \centering
    \includegraphics[width=.49\textwidth]{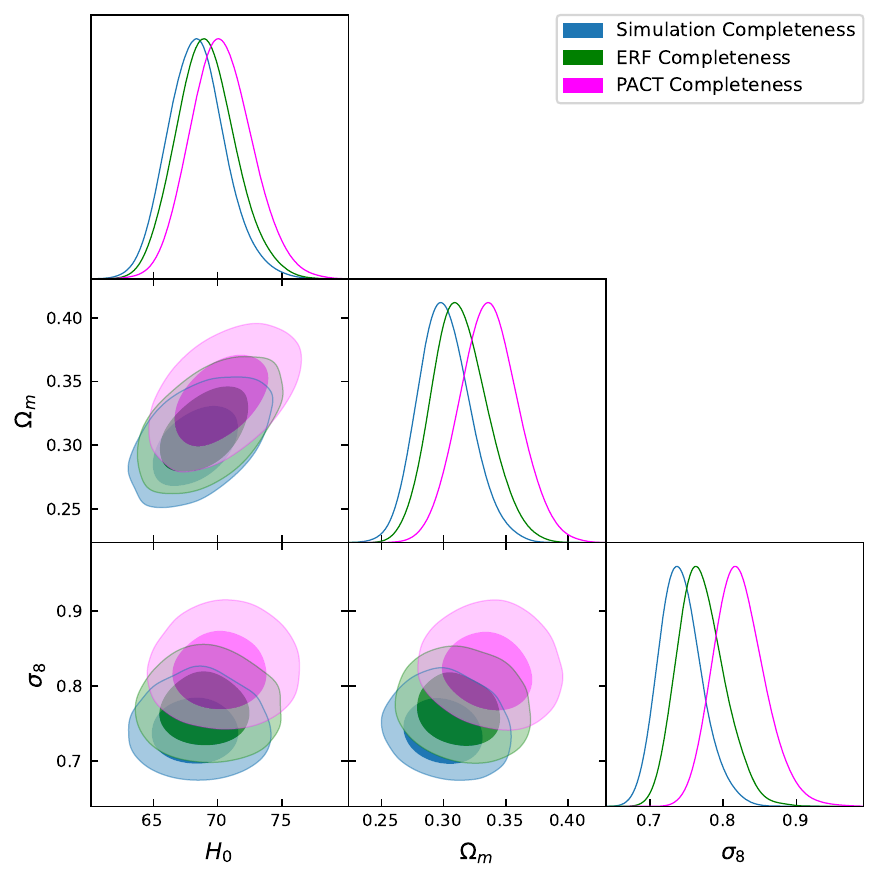}
    \caption{Constraints on cosmological parameters ($H_0$, $\Omega_m$, $\sigma_8$) from the \Planck PSZ2 cluster number counts and BAO data, with three different completeness functions: the ``standard'' ERF completeness (same as in \citealp{PLANCK2016-PSZ2_cosmology}), and two ``fitted'' versions obtained rescaling the noise per patch by a constant, to reproduce the completeness derived from the simulation clusters and the PACT \citep{Pointecouteau2021-profile-PACT} profile clusters. The filled contours represent the $68\%$ and $95\%$ confidence regions.}
    \label{fig:cosmo_params_newCompleteness}
\end{figure}

\begin{figure}
    \centering
    \includegraphics[width=.4\textwidth]{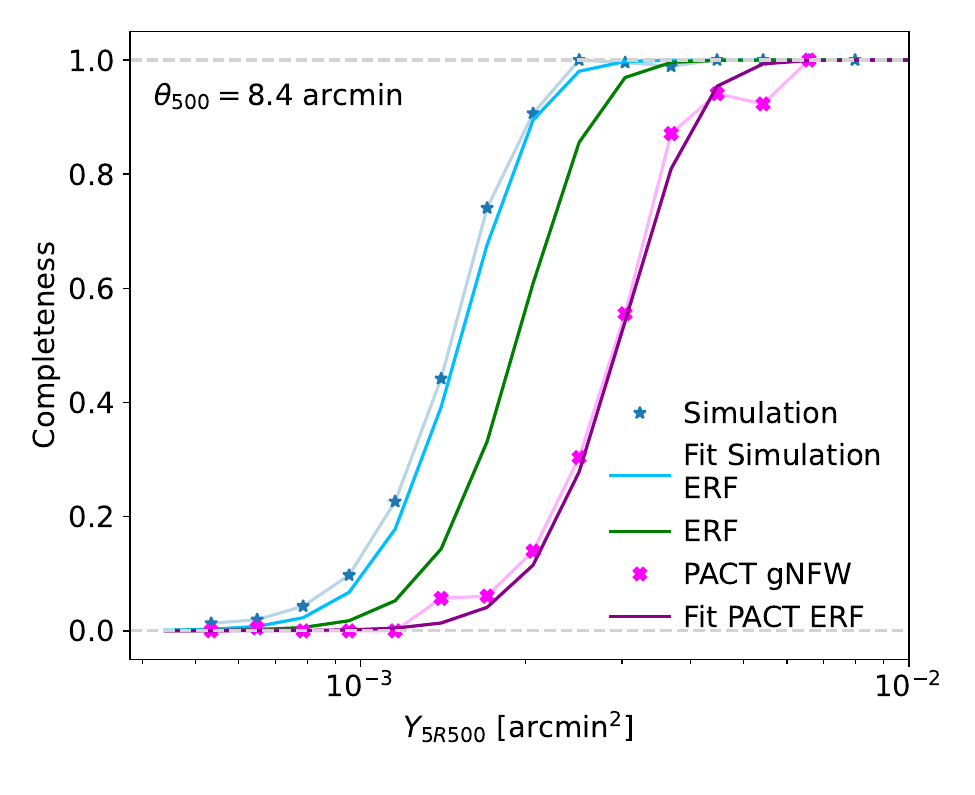}
    \caption{Comparison of the three completeness functions used in the cosmological analyses with the ones computed via Monte-Carlo injection of simulation images and spherical images with the \cite{Pointecouteau2021-profile-PACT} profile.}
    \label{fig:fitted_ERF_completeness}
\end{figure}

Finally, we test the impact of changing the completeness function in the cosmological analysis. We choose two completeness examples from the different sets we obtained, one above and one below the ERF estimate: the simulation completeness and the PACT gNFW completeness. By changing only the completeness function, and not the other elements of the analysis (like the scaling relations and mass bias), we are actually testing a rather extreme scenario, in which the true cluster profile is different not only from the one assumed in the MMF templates, but also in the scaling relations calibration and mass bias estimations. So, the results we obtain here are to be taken as indicative of a trend of impact, rather than for the actual values.

To simplify the analysis, we choose to approximate the completeness obtained from Monte-Carlo injection with an adapted version of the ERF completeness.
We use a simple prescription to approximate the completeness: we modify the functional form of the ERF completeness in Eq. \ref{eq:erf_completeness} by adding a free multiplicative parameter in front of the per-patch noise of the original \Planck maps $\sigma_{Yi}^{\mathrm{new}}(\theta_{500}) = a_\mathrm{fit}\, \sigma_{Yi}(\theta_{500})$, and fit it to the two completeness obtained from Monte-Carlo injection. The fitted value for the simulation completeness is $a_\mathrm{fit}^\mathrm{sim}=1.27$, and the one for the PACT completeness is $a_\mathrm{fit}^\mathrm{PACT}=0.65$. We show an example of the fitted ERF completeness compared to the Monte-Carlo obtained ones in Fig. \ref{fig:fitted_ERF_completeness}. Up to now we used a threshold in S/N of 4.5 to compute the completeness, but the \Planck cosmological sample has a threshold of S/N $>6$. Anyway, we checked that the values of the noise shift we computed are the same also with higher S/N threshold, so we can simply set $q=6$ in Eq. \ref{eq:erf_completeness} to get the ERF completeness for the cosmological sample.

We perform a cosmological analysis of cluster number counts, following \cite{PLANCK2016-PSZ2_cosmology}. We use the \Planck PSZ2 MMF3 cosmological cluster sample \citep{PLANCK2015-PSZ2}, adding observational constraints from Big Bang nucleosyntesis \citep[BBN][]{Steigman2008-BBN}, baryon acoustic oscillations (BAO) measurements from SDSS-III DR12 \citep{Alam2017-BAO_SDSS_BOSS_DR12}, and a prior on $n_s$ from \cite{PLANCK2014-Cosmology}. For the mass-observable scaling relations and mass bias parameters we use the same priors as the baseline analysis in \cite{PLANCK2016-PSZ2_cosmology}. The constraints on the cosmological parameters are obtained with a Markov-Chain Monte-Carlo (MCMC), implemented in the code \texttt{CosmoMC}\footnote{\url{https://cosmologist.info/cosmomc/}} \citep{Lewis2002-CosmoMC}.
We perform the analysis using the original ERF completeness and the approximated Simulation and PACT completeness described above. The result of the three analyses are compared in Fig. \ref{fig:cosmo_params_newCompleteness}, for $H_0$, $\Omega_m$ and $\sigma_8$, and we list the constraints in Table \ref{tab:cosmo_params_constraints}.

From these results, one can notice how the change in the completeness moves the constraints on $\Omega_m$ and $\sigma_8$. In particular, a higher completeness with respect to the ERF (in this case the Simulation one) favors lower values of the two cosmological parameters, while a lower completeness (the PACT one) prefers higher values. For $\Omega_m$, the shift amounts to about $0.6\sigma$ for the Simulation completeness, and about $1.1\sigma$ for the PACT one. For $\sigma_8$, the two shifts are $0.9\sigma$ and $1.8\sigma$, respectively. It is interesting to notice how the direction of the shift in the $(\Omega_m, \sigma_8)$ plane is the same as moving the prior on the mass bias, as shown in Fig. 7 of \cite{PLANCK2016-PSZ2_cosmology}.

\begin{table}
    \centering
    \caption{Constraints on cosmological parameters obtained from the analysis of \Planck PSZ2 cluster number counts and BAO, using three different completeness functions described in the text.}
    \label{tab:cosmo_params_constraints}
    \begingroup
    \renewcommand{\arraystretch}{1.5}
    \begin{tabular}{lrrr}
	\toprule
	 & Simulation & ERF & PACT \\
	 & Completeness & Completeness & Completeness \\
	\midrule
	$\mathbf{H_0}$ & $68.4^{+2.0}_{-2.4}$ & $69.1^{+2.1}_{-2.4}$ & $70.3^{+2.2}_{-2.5}$ \\ 
	$\mathbf{\Omega_m}$ & $0.301^{+0.019}_{-0.023}$ & $0.314^{+0.020}_{-0.024}$ & $0.337\pm 0.024$ \\ 
	$\mathbf{\sigma_8}$ & $0.742^{+0.025}_{-0.033}$ & $0.770^{+0.027}_{-0.036}$ & $0.823^{+0.030}_{-0.039}$ \\ 
	\bottomrule
\end{tabular}

    \endgroup
\end{table}

\section{Discussion}\label{sec:discussion}

A clear limit of our set of simulation images is the restricted range of scales we are able to analyse, approximately $[4-10]$ arcmin, compared to the extent of the \Planck catalog. This is due to the limits of the simulation: first, the limited volume of the simulation doesn't allow for many high mass clusters to form. This becomes a limiting factor at high redshift, because there aren't enough high mass clusters to sample the high signal tail of the completeness. This translates to a lower limit in $\theta_{500}$. The second limit is related to the spacing of the simulation snapshots in redshift, which is about 0.01 at low $z$. When translating the cluster distribution from the mass-redshift plane to the ($\theta_{500}$, $Y_{5R500}$) one, we see large empty regions appear diagonally at high $\theta_{500}$, thus preventing a good coverage of the completeness in that region, and effectively imposing a higher limit on the cluster scale.

This limitation prevents us from probing the completeness in the extremes of the \Planck cluster sample, and the possible impacts of using cluster images that differ from the detection template in these extreme contexts. For example, at large cluster scales there could be more important effects due to asphericity of clusters, or with broader range of scales one could be able to test more in detail the scale dependence of the completeness beyond the ERF model.

Another aspect that might be deemed too simplistic is the use of the same pressure profile in all the injected spherical images for each of the sets we use. In this way the cluster signals we inject are all just scaled versions of each other. This is obviously unrealistic, and neglects the scatter around the mean profile which is present in the data. The main reason is that, given that we already incorporate a realistic treatment with the Simulation set, we treat the sets of spherically symmetric images as idealised test cases, and use them to understand the response of the detection algorithm to different profile shapes, without the pretence of being realistic. Moreover, the impact of profile scatter between spherical images has been already detailed in \cite{PLANCK2014-PSZ, PLANCK2015-PSZ2}, that found a widening effect in the completeness drop-off region.

A third possible criticism to the work presented so far is the use of just one simulation. Indeed, the specific shape and values of the completeness function we find for the simulation set are certainly simulation-dependent to an extent. Nonetheless, we believe that the main result of this work is the apparent dependence of the completeness on more cluster characteristics than just the ones of the ERF estimate (i.e. cluster size and total integrated signal), notably the steepness of the Compton-$y$ profile, which we show having an important impact on the completeness.

\section{Conclusion}\label{sec:conclusion}

In this work we have analysed the completeness function for galaxy clusters detected in SZ by \Planck with the MMF algorithm, focusing on the case where the cluster model assumed in the detection method differs from the ``true'' one, both in terms of shape and pressure profile.
We have done so via Monte-Carlo injection of SZ signal from simulated clusters into cleaned \Planck sky maps. Comparing the injected sources with the ones detected by the MMF algorithm, we computed the completeness in bins of injected $(Y_{5R500}, \theta_{500})$ (i.e. clusters' total SZ signal and radius). 
We used a set of realistic cluster images from the IllustrisTNG simulation, as well as images of spherically symmetric cluster generated from different pressure profiles.
We found that the completeness computed with the simulation images is generally higher than the ERF completeness estimate, which is based on the assumption of Gaussian noise in the SZ signal. For the same cluster set we also reported biased estimations of the clusters' observables, with underestimation of the cluster size, $\theta_{500}$, and overestimation of the total signal $Y_{5R500}$. We believe that these effects are mainly related to the fact that the average Compton-$y$ profile of the simulated clusters is different, and in particular steeper, than the one used as template for the detection. To confirm this explanation, we repeated the same procedure with different sets of images, generated from spherically symmetric gNFW pressure profiles with different sets of parameters. We found that when using the parameters from \cite{Arnaud2010} (the same of the detection template), the completeness is consistent with the ERF estimate. But when changing the profile parameters, we found that a flatter profile \citep[like the ones of][]{PLANCK2013-profile, Pointecouteau2021-profile-PACT} produces completeness functions that are lower than the ERF, while steeper profiles \citep[for example][$z<0.35$]{Tramonte2023-profile-stacked} have generally higher completeness than the ERF.

We also tested the impact of the asymmetric cluster morphology of our simulated set on the completeness. Comparing the most and least spherical subsets, we found that the effect of the asphericity is generally small, especially below the \Planck beam size, but there is a tendency of the ``more elliptical'' subset to be systematically less complete than the ``more spherical'' one. The effect tends to increase the more the clusters are resolved, thus for larger $\theta_{500}$, at least in the range covered by our cluster set.

Finally, we studied the effect of modifying the completeness function in the cosmological analysis with galaxy cluster number counts. We found, taking two completeness examples that deviate substantially from the ERF estimate, that the impact on the cosmological parameters $(\Omega_m, \sigma_8)$ is a shift in the constraints of roughly one $\sigma$, with higher completeness favoring lower values of the cosmological parameters, and vice versa. We recall that what we tested here is an extreme scenario, in which the true cluster are different for the ones used to constrain the detection template, scaling relations, mass bias, etc. What is interesting to remark, anyway, is that the shift due to changing the completeness goes in the same direction as the one due to the mass bias, and therefore the two effects should probably be taken into account together. This work highlights the fact that any bias and uncertainty on the completeness should be propagated into the galaxy cluster number count likelihood for robust and accurate analyses.

\begin{acknowledgements}

The authors acknowledge the fruitful discussions and comments from Nabila Aghanim and the other members of the Cosmology team at IAS.
SG acknowledges financial support from the Ecole Doctorale d’Astronomie et d’Astrophysique d’Ile-de-France (ED AAIF). This research has made use of the computation facility of the Integrated Data and Operation Center (IDOC, \url{https://idoc.ias.u-psud.fr}) at the Institut d’Astrophysique Spatiale (IAS), as well as the SZ-Cluster Database (\url{https://szdb.osups.universite-paris-saclay.fr}). This project was carried out using the Python libraries \texttt{matplotlib} \citep{Hunter2007-matplotlib}, \texttt{numpy} \citep{harris2020-numpy}, \texttt{Astropy} \citep{astropy2013, astropy2018, astropy2022}. 

\end{acknowledgements}

\bibliographystyle{aa}
\bibliography{references}

\begin{thebibliography}{61}
\expandafter\ifx\csname natexlab\endcsname\relax\def\natexlab#1{#1}\fi

\bibitem[{Abbott {et~al.}(2020)Abbott, Aguena, Alarcon, Allam, Allen, Annis,
  Avila, Bacon, Bechtol, Bermeo, Bernstein, Bertin, Bhargava, Bocquet, Brooks,
  Brout, Buckley-Geer, Burke, Carnero~Rosell, Carrasco~Kind, Carretero,
  Castander, Cawthon, Chang, Chen, Choi, Costanzi, Crocce, da~Costa, Davis,
  De~Vicente, DeRose, Desai, Diehl, Dietrich, Dodelson, Doel, Drlica-Wagner,
  Eckert, Eifler, Elvin-Poole, Estrada, Everett, Evrard, Farahi, Ferrero,
  Flaugher, Fosalba, Frieman, Garc\'{\i}a-Bellido, Gatti, Gaztanaga, Gerdes,
  Giannantonio, Giles, Grandis, Gruen, Gruendl, Gschwend, Gutierrez, Hartley,
  Hinton, Hollowood, Honscheid, Hoyle, Huterer, James, Jarvis, Jeltema,
  Johnson, Johnson, Kent, Krause, Kron, Kuehn, Kuropatkin, Lahav, Li, Lidman,
  Lima, Lin, MacCrann, Maia, Mantz, Marshall, Martini, Mayers, Melchior,
  Mena-Fern\'andez, Menanteau, Miquel, Mohr, Nichol, Nord, Ogando, Palmese,
  Paz-Chinch\'on, Plazas, Prat, Rau, Romer, Roodman, Rooney, Rozo, Rykoff,
  Sako, Samuroff, S\'anchez, Sanchez, Saro, Scarpine, Schubnell, Scolnic,
  Serrano, Sevilla-Noarbe, Sheldon, Smith, Smith, Suchyta, Swanson, Tarle,
  Thomas, To, Troxel, Tucker, Varga, von~der Linden, Walker, Wechsler, Weller,
  Wilkinson, Wu, Yanny, Zhang, Zhang, \& Zuntz}]{Abbott2020-DES}
Abbott, T. M.~C., Aguena, M., Alarcon, A., {et~al.} 2020, Phys. Rev. D, 102,
  023509

\bibitem[{{Aghanim} {et~al.}(2019){Aghanim}, {Douspis}, {Hurier}, {Crichton},
  {Diego}, {Hasselfield}, {Macias-Perez}, {Marriage}, {Pointecouteau},
  {Remazeilles}, \& {Soubri{\'e}}}]{Aghanim2019-PACT}
{Aghanim}, N., {Douspis}, M., {Hurier}, G., {et~al.} 2019, \aap, 632, A47

\bibitem[{{Alam} {et~al.}(2017){Alam}, {Ata}, {Bailey}, {Beutler}, {Bizyaev},
  {Blazek}, {Bolton}, {Brownstein}, {Burden}, {Chuang}, {Comparat}, {Cuesta},
  {Dawson}, {Eisenstein}, {Escoffier}, {Gil-Mar{\'\i}n}, {Grieb}, {Hand}, {Ho},
  {Kinemuchi}, {Kirkby}, {Kitaura}, {Malanushenko}, {Malanushenko}, {Maraston},
  {McBride}, {Nichol}, {Olmstead}, {Oravetz}, {Padmanabhan},
  {Palanque-Delabrouille}, {Pan}, {Pellejero-Ibanez}, {Percival}, {Petitjean},
  {Prada}, {Price-Whelan}, {Reid}, {Rodr{\'\i}guez-Torres}, {Roe}, {Ross},
  {Ross}, {Rossi}, {Rubi{\~n}o-Mart{\'\i}n}, {Saito}, {Salazar-Albornoz},
  {Samushia}, {S{\'a}nchez}, {Satpathy}, {Schlegel}, {Schneider},
  {Sc{\'o}ccola}, {Seo}, {Sheldon}, {Simmons}, {Slosar}, {Strauss}, {Swanson},
  {Thomas}, {Tinker}, {Tojeiro}, {Maga{\~n}a}, {Vazquez}, {Verde}, {Wake},
  {Wang}, {Weinberg}, {White}, {Wood-Vasey}, {Y{\`e}che}, {Zehavi}, {Zhai}, \&
  {Zhao}}]{Alam2017-BAO_SDSS_BOSS_DR12}
{Alam}, S., {Ata}, M., {Bailey}, S., {et~al.} 2017, \mnras, 470, 2617

\bibitem[{{Allen} {et~al.}(2011){Allen}, {Evrard}, \& {Mantz}}]{Allen2011}
{Allen}, S.~W., {Evrard}, A.~E., \& {Mantz}, A.~B. 2011, \araa, 49, 409

\bibitem[{{Arnaud} {et~al.}(2010){Arnaud}, {Pratt}, {Piffaretti},
  {B{\"o}hringer}, {Croston}, \& {Pointecouteau}}]{Arnaud2010}
{Arnaud}, M., {Pratt}, G.~W., {Piffaretti}, R., {et~al.} 2010, \aap, 517, A92

\bibitem[{{Astropy Collaboration} {et~al.}(2022){Astropy Collaboration},
  {Price-Whelan}, {Lim}, {Earl}, {Starkman}, {Bradley}, {Shupe}, {Patil},
  {Corrales}, {Brasseur}, {N{"o}the}, {Donath}, {Tollerud}, {Morris},
  {Ginsburg}, {Vaher}, {Weaver}, {Tocknell}, {Jamieson}, {van Kerkwijk},
  {Robitaille}, {Merry}, {Bachetti}, {G{"u}nther}, {Aldcroft},
  {Alvarado-Montes}, {Archibald}, {B{'o}di}, {Bapat}, {Barentsen}, {Baz{'a}n},
  {Biswas}, {Boquien}, {Burke}, {Cara}, {Cara}, {Conroy}, {Conseil}, {Craig},
  {Cross}, {Cruz}, {D'Eugenio}, {Dencheva}, {Devillepoix}, {Dietrich},
  {Eigenbrot}, {Erben}, {Ferreira}, {Foreman-Mackey}, {Fox}, {Freij}, {Garg},
  {Geda}, {Glattly}, {Gondhalekar}, {Gordon}, {Grant}, {Greenfield}, {Groener},
  {Guest}, {Gurovich}, {Handberg}, {Hart}, {Hatfield-Dodds}, {Homeier},
  {Hosseinzadeh}, {Jenness}, {Jones}, {Joseph}, {Kalmbach}, {Karamehmetoglu},
  {Ka{l}uszy{'n}ski}, {Kelley}, {Kern}, {Kerzendorf}, {Koch}, {Kulumani},
  {Lee}, {Ly}, {Ma}, {MacBride}, {Maljaars}, {Muna}, {Murphy}, {Norman},
  {O'Steen}, {Oman}, {Pacifici}, {Pascual}, {Pascual-Granado}, {Patil},
  {Perren}, {Pickering}, {Rastogi}, {Roulston}, {Ryan}, {Rykoff}, {Sabater},
  {Sakurikar}, {Salgado}, {Sanghi}, {Saunders}, {Savchenko}, {Schwardt},
  {Seifert-Eckert}, {Shih}, {Jain}, {Shukla}, {Sick}, {Simpson},
  {Singanamalla}, {Singer}, {Singhal}, {Sinha}, {Sip{H{o}}cz}, {Spitler},
  {Stansby}, {Streicher}, {{{S}}umak}, {Swinbank}, {Taranu}, {Tewary},
  {Tremblay}, {Val-Borro}, {Van Kooten}, {Vasovi{'c}}, {Verma}, {de Miranda
  Cardoso}, {Williams}, {Wilson}, {Winkel}, {Wood-Vasey}, {Xue}, {Yoachim},
  {Zhang}, {Zonca}, \& {Astropy Project Contributors}}]{astropy2022}
{Astropy Collaboration}, {Price-Whelan}, A.~M., {Lim}, P.~L., {et~al.} 2022,
  apj, 935, 167

\bibitem[{{Astropy Collaboration} {et~al.}(2018){Astropy Collaboration},
  {Price-Whelan}, {Sip{\H{o}}cz}, {G{\"u}nther}, {Lim}, {Crawford}, {Conseil},
  {Shupe}, {Craig}, {Dencheva}, {Ginsburg}, {Vand erPlas}, {Bradley},
  {P{\'e}rez-Su{\'a}rez}, {de Val-Borro}, {Aldcroft}, {Cruz}, {Robitaille},
  {Tollerud}, {Ardelean}, {Babej}, {Bach}, {Bachetti}, {Bakanov}, {Bamford},
  {Barentsen}, {Barmby}, {Baumbach}, {Berry}, {Biscani}, {Boquien}, {Bostroem},
  {Bouma}, {Brammer}, {Bray}, {Breytenbach}, {Buddelmeijer}, {Burke},
  {Calderone}, {Cano Rodr{\'\i}guez}, {Cara}, {Cardoso}, {Cheedella}, {Copin},
  {Corrales}, {Crichton}, {D'Avella}, {Deil}, {Depagne}, {Dietrich}, {Donath},
  {Droettboom}, {Earl}, {Erben}, {Fabbro}, {Ferreira}, {Finethy}, {Fox},
  {Garrison}, {Gibbons}, {Goldstein}, {Gommers}, {Greco}, {Greenfield},
  {Groener}, {Grollier}, {Hagen}, {Hirst}, {Homeier}, {Horton}, {Hosseinzadeh},
  {Hu}, {Hunkeler}, {Ivezi{\'c}}, {Jain}, {Jenness}, {Kanarek}, {Kendrew},
  {Kern}, {Kerzendorf}, {Khvalko}, {King}, {Kirkby}, {Kulkarni}, {Kumar},
  {Lee}, {Lenz}, {Littlefair}, {Ma}, {Macleod}, {Mastropietro}, {McCully},
  {Montagnac}, {Morris}, {Mueller}, {Mumford}, {Muna}, {Murphy}, {Nelson},
  {Nguyen}, {Ninan}, {N{\"o}the}, {Ogaz}, {Oh}, {Parejko}, {Parley}, {Pascual},
  {Patil}, {Patil}, {Plunkett}, {Prochaska}, {Rastogi}, {Reddy Janga},
  {Sabater}, {Sakurikar}, {Seifert}, {Sherbert}, {Sherwood-Taylor}, {Shih},
  {Sick}, {Silbiger}, {Singanamalla}, {Singer}, {Sladen}, {Sooley},
  {Sornarajah}, {Streicher}, {Teuben}, {Thomas}, {Tremblay}, {Turner},
  {Terr{\'o}n}, {van Kerkwijk}, {de la Vega}, {Watkins}, {Weaver}, {Whitmore},
  {Woillez}, {Zabalza}, \& {Astropy Contributors}}]{astropy2018}
{Astropy Collaboration}, {Price-Whelan}, A.~M., {Sip{\H{o}}cz}, B.~M., {et~al.}
  2018, \aj, 156, 123

\bibitem[{{Astropy Collaboration} {et~al.}(2013){Astropy Collaboration},
  {Robitaille}, {Tollerud}, {Greenfield}, {Droettboom}, {Bray}, {Aldcroft},
  {Davis}, {Ginsburg}, {Price-Whelan}, {Kerzendorf}, {Conley}, {Crighton},
  {Barbary}, {Muna}, {Ferguson}, {Grollier}, {Parikh}, {Nair}, {Unther},
  {Deil}, {Woillez}, {Conseil}, {Kramer}, {Turner}, {Singer}, {Fox}, {Weaver},
  {Zabalza}, {Edwards}, {Azalee Bostroem}, {Burke}, {Casey}, {Crawford},
  {Dencheva}, {Ely}, {Jenness}, {Labrie}, {Lim}, {Pierfederici}, {Pontzen},
  {Ptak}, {Refsdal}, {Servillat}, \& {Streicher}}]{astropy2013}
{Astropy Collaboration}, {Robitaille}, T.~P., {Tollerud}, E.~J., {et~al.} 2013,
  \aap, 558, A33

\bibitem[{{Bleem} {et~al.}(2015){Bleem}, {Stalder}, {de Haan}, {Aird}, {Allen},
  {Applegate}, {Ashby}, {Bautz}, {Bayliss}, {Benson}, {Bocquet}, {Brodwin},
  {Carlstrom}, {Chang}, {Chiu}, {Cho}, {Clocchiatti}, {Crawford}, {Crites},
  {Desai}, {Dietrich}, {Dobbs}, {Foley}, {Forman}, {George}, {Gladders},
  {Gonzalez}, {Halverson}, {Hennig}, {Hoekstra}, {Holder}, {Holzapfel},
  {Hrubes}, {Jones}, {Keisler}, {Knox}, {Lee}, {Leitch}, {Liu}, {Lueker},
  {Luong-Van}, {Mantz}, {Marrone}, {McDonald}, {McMahon}, {Meyer}, {Mocanu},
  {Mohr}, {Murray}, {Padin}, {Pryke}, {Reichardt}, {Rest}, {Ruel}, {Ruhl},
  {Saliwanchik}, {Saro}, {Sayre}, {Schaffer}, {Schrabback}, {Shirokoff},
  {Song}, {Spieler}, {Stanford}, {Staniszewski}, {Stark}, {Story}, {Stubbs},
  {Vanderlinde}, {Vieira}, {Vikhlinin}, {Williamson}, {Zahn}, \&
  {Zenteno}}]{Bleem2015-SPT-SZcatalog}
{Bleem}, L.~E., {Stalder}, B., {de Haan}, T., {et~al.} 2015, \apjs, 216, 27

\bibitem[{{Bocquet} {et~al.}(2019){Bocquet}, {Dietrich}, {Schrabback}, {Bleem},
  {Klein}, {Allen}, {Applegate}, {Ashby}, {Bautz}, {Bayliss}, {Benson},
  {Brodwin}, {Bulbul}, {Canning}, {Capasso}, {Carlstrom}, {Chang}, {Chiu},
  {Cho}, {Clocchiatti}, {Crawford}, {Crites}, {de Haan}, {Desai}, {Dobbs},
  {Foley}, {Forman}, {Garmire}, {George}, {Gladders}, {Gonzalez}, {Grandis},
  {Gupta}, {Halverson}, {Hlavacek-Larrondo}, {Hoekstra}, {Holder}, {Holzapfel},
  {Hou}, {Hrubes}, {Huang}, {Jones}, {Khullar}, {Knox}, {Kraft}, {Lee}, {von
  der Linden}, {Luong-Van}, {Mantz}, {Marrone}, {McDonald}, {McMahon}, {Meyer},
  {Mocanu}, {Mohr}, {Morris}, {Padin}, {Patil}, {Pryke}, {Rapetti},
  {Reichardt}, {Rest}, {Ruhl}, {Saliwanchik}, {Saro}, {Sayre}, {Schaffer},
  {Shirokoff}, {Stalder}, {Stanford}, {Staniszewski}, {Stark}, {Story},
  {Strazzullo}, {Stubbs}, {Vanderlinde}, {Vieira}, {Vikhlinin}, {Williamson},
  \& {Zenteno}}]{Bocquet2019-SPT_SZ_cluster_cosmology}
{Bocquet}, S., {Dietrich}, J.~P., {Schrabback}, T., {et~al.} 2019, \apj, 878,
  55

\bibitem[{{Costanzi} {et~al.}(2021){Costanzi}, {Saro}, {Bocquet}, {Abbott},
  {Aguena}, {Allam}, {Amara}, {Annis}, {Avila}, {Bacon}, {Benson}, {Bhargava},
  {Brooks}, {Buckley-Geer}, {Burke}, {Carnero Rosell}, {Carrasco Kind},
  {Carretero}, {Choi}, {da Costa}, {Pereira}, {De Vicente}, {Desai}, {Diehl},
  {Dietrich}, {Doel}, {Eifler}, {Everett}, {Ferrero}, {Fert{\'e}}, {Flaugher},
  {Fosalba}, {Frieman}, {Garc{\'\i}a-Bellido}, {Gaztanaga}, {Gerdes},
  {Giannantonio}, {Giles}, {Grandis}, {Gruen}, {Gruendl}, {Gupta}, {Gutierrez},
  {Hartley}, {Hinton}, {Hollowood}, {Honscheid}, {James}, {Jeltema}, {Krause},
  {Kuehn}, {Kuropatkin}, {Lahav}, {Lima}, {MacCrann}, {Maia}, {Marshall},
  {Menanteau}, {Miquel}, {Mohr}, {Morgan}, {Myles}, {Ogando}, {Palmese},
  {Paz-Chinch{\'o}n}, {Plazas}, {Rapetti}, {Reichardt}, {Romer}, {Roodman},
  {Ruppin}, {Salvati}, {Samuroff}, {Sanchez}, {Scarpine}, {Serrano},
  {Sevilla-Noarbe}, {Singh}, {Smith}, {Soares-Santos}, {Stark}, {Suchyta},
  {Swanson}, {Tarle}, {Thomas}, {To}, {Tucker}, {Varga}, {Wechsler}, {Zhang},
  {DES}, \& {SPT Collaborations}}]{Costanzi2021-DESY1_SPT_cluster_cosmology}
{Costanzi}, M., {Saro}, A., {Bocquet}, S., {et~al.} 2021, \prd, 103, 043522

\bibitem[{{Davis} {et~al.}(1985){Davis}, {Efstathiou}, {Frenk}, \&
  {White}}]{Davis1985}
{Davis}, M., {Efstathiou}, G., {Frenk}, C.~S., \& {White}, S.~D.~M. 1985, \apj,
  292, 371

\bibitem[{{G{\'o}rski} {et~al.}(2005){G{\'o}rski}, {Hivon}, {Banday},
  {Wandelt}, {Hansen}, {Reinecke}, \& {Bartelmann}}]{Gorski2005-HEALPix}
{G{\'o}rski}, K.~M., {Hivon}, E., {Banday}, A.~J., {et~al.} 2005, \apj, 622,
  759

\bibitem[{{Gouin} {et~al.}(2020){Gouin}, {Aghanim}, {Bonjean}, \&
  {Douspis}}]{Gouin2019}
{Gouin}, C., {Aghanim}, N., {Bonjean}, V., \& {Douspis}, M. 2020, \aap, 635,
  A195

\bibitem[{{Gouin} {et~al.}(2022){Gouin}, {Gallo}, \& {Aghanim}}]{Gouin2022}
{Gouin}, C., {Gallo}, S., \& {Aghanim}, N. 2022, \aap, 664, A198

\bibitem[{Harris {et~al.}(2020)Harris, Millman, van~der Walt, Gommers,
  Virtanen, Cournapeau, Wieser, Taylor, Berg, Smith, Kern, Picus, Hoyer, van
  Kerkwijk, Brett, Haldane, del R{\'{i}}o, Wiebe, Peterson,
  G{\'{e}}rard-Marchant, Sheppard, Reddy, Weckesser, Abbasi, Gohlke, \&
  Oliphant}]{harris2020-numpy}
Harris, C.~R., Millman, K.~J., van~der Walt, S.~J., {et~al.} 2020, Nature, 585,
  357

\bibitem[{{Henry}(1997)}]{Henry1997}
{Henry}, J.~P. 1997, \apjl, 489, L1

\bibitem[{{Herranz} {et~al.}(2002){Herranz}, {Sanz}, {Barreiro}, \&
  {Mart{\'\i}nez-Gonz{\'a}lez}}]{Herranz2002-MMF}
{Herranz}, D., {Sanz}, J.~L., {Barreiro}, R.~B., \&
  {Mart{\'\i}nez-Gonz{\'a}lez}, E. 2002, \apj, 580, 610

\bibitem[{{Hilton} {et~al.}(2021){Hilton}, {Sif{\'o}n}, {Naess},
  {Madhavacheril}, {Oguri}, {Rozo}, {Rykoff}, {Abbott}, {Adhikari}, {Aguena},
  {Aiola}, {Allam}, {Amodeo}, {Amon}, {Annis}, {Ansarinejad}, {Aros-Bunster},
  {Austermann}, {Avila}, {Bacon}, {Battaglia}, {Beall}, {Becker}, {Bernstein},
  {Bertin}, {Bhandarkar}, {Bhargava}, {Bond}, {Brooks}, {Burke}, {Calabrese},
  {Carrasco Kind}, {Carretero}, {Choi}, {Choi}, {Conselice}, {da Costa},
  {Costanzi}, {Crichton}, {Crowley}, {D{\"u}nner}, {Denison}, {Devlin},
  {Dicker}, {Diehl}, {Dietrich}, {Doel}, {Duff}, {Duivenvoorden}, {Dunkley},
  {Everett}, {Ferraro}, {Ferrero}, {Fert{\'e}}, {Flaugher}, {Frieman},
  {Gallardo}, {Garc{\'\i}a-Bellido}, {Gaztanaga}, {Gerdes}, {Giles}, {Golec},
  {Gralla}, {Grandis}, {Gruen}, {Gruendl}, {Gschwend}, {Gutierrez}, {Han},
  {Hartley}, {Hasselfield}, {Hill}, {Hilton}, {Hincks}, {Hinton}, {Ho},
  {Honscheid}, {Hoyle}, {Hubmayr}, {Huffenberger}, {Hughes}, {Jaelani}, {Jain},
  {James}, {Jeltema}, {Kent}, {Knowles}, {Koopman}, {Kuehn}, {Lahav}, {Lima},
  {Lin}, {Lokken}, {Loubser}, {MacCrann}, {Maia}, {Marriage}, {Martin},
  {McMahon}, {Melchior}, {Menanteau}, {Miquel}, {Miyatake}, {Moodley},
  {Morgan}, {Mroczkowski}, {Nati}, {Newburgh}, {Niemack}, {Nishizawa},
  {Ogando}, {Orlowski-Scherer}, {Page}, {Palmese}, {Partridge},
  {Paz-Chinch{\'o}n}, {Phakathi}, {Plazas}, {Robertson}, {Romer}, {Carnero
  Rosell}, {Salatino}, {Sanchez}, {Schaan}, {Schillaci}, {Sehgal}, {Serrano},
  {Shin}, {Simon}, {Smith}, {Soares-Santos}, {Spergel}, {Staggs}, {Storer},
  {Suchyta}, {Swanson}, {Tarle}, {Thomas}, {To}, {Trac}, {Ullom}, {Vale}, {Van
  Lanen}, {Vavagiakis}, {De Vicente}, {Wilkinson}, {Wollack}, {Xu}, \&
  {Zhang}}]{Hilton2021-ACT-SZcatalog}
{Hilton}, M., {Sif{\'o}n}, C., {Naess}, S., {et~al.} 2021, \apjs, 253, 3

\bibitem[{Hunter(2007)}]{Hunter2007-matplotlib}
Hunter, J.~D. 2007, Computing in Science \& Engineering, 9, 90

\bibitem[{{Kravtsov} \&
  {Borgani}(2012)}]{Kravtsov&Borgani2012-cluster_formation_review}
{Kravtsov}, A.~V. \& {Borgani}, S. 2012, \araa, 50, 353

\bibitem[{{Lewis} \& {Bridle}(2002)}]{Lewis2002-CosmoMC}
{Lewis}, A. \& {Bridle}, S. 2002, \prd, 66, 103511

\bibitem[{{Limousin} {et~al.}(2013){Limousin}, {Morandi}, {Sereno},
  {Meneghetti}, {Ettori}, {Bartelmann}, \& {Verdugo}}]{Limousin2013}
{Limousin}, M., {Morandi}, A., {Sereno}, M., {et~al.} 2013, \ssr, 177, 155

\bibitem[{{Liu} {et~al.}(2022){Liu}, {Bulbul}, {Ghirardini}, {Liu}, {Klein},
  {Clerc}, {{\"O}zsoy}, {Ramos-Ceja}, {Pacaud}, {Comparat}, {Okabe}, {Bahar},
  {Biffi}, {Brunner}, {Br{\"u}ggen}, {Buchner}, {Ider Chitham}, {Chiu},
  {Dolag}, {Gatuzz}, {Gonzalez}, {Hoang}, {Lamer}, {Merloni}, {Nandra},
  {Oguri}, {Ota}, {Predehl}, {Reiprich}, {Salvato}, {Schrabback}, {Sanders},
  {Seppi}, \& {Thibaud}}]{Liu2022-eROSITA-clusters}
{Liu}, A., {Bulbul}, E., {Ghirardini}, V., {et~al.} 2022, \aap, 661, A2

\bibitem[{{Marinacci} {et~al.}(2018){Marinacci}, {Vogelsberger}, {Pakmor},
  {Torrey}, {Springel}, {Hernquist}, {Nelson}, {Weinberger}, {Pillepich},
  {Naiman}, \& {Genel}}]{TNG(f)}
{Marinacci}, F., {Vogelsberger}, M., {Pakmor}, R., {et~al.} 2018, \mnras, 480,
  5113

\bibitem[{{Melin} {et~al.}(2005){Melin}, {Bartlett}, \&
  {Delabrouille}}]{Melin2005-SZ_selection_function}
{Melin}, J.~B., {Bartlett}, J.~G., \& {Delabrouille}, J. 2005, \aap, 429, 417

\bibitem[{{Melin} {et~al.}(2006){Melin}, {Bartlett}, \&
  {Delabrouille}}]{Melin2006-MMF}
{Melin}, J.~B., {Bartlett}, J.~G., \& {Delabrouille}, J. 2006, \aap, 459, 341

\bibitem[{{Nagai} {et~al.}(2007){Nagai}, {Kravtsov}, \&
  {Vikhlinin}}]{Nagai2007-gNFW}
{Nagai}, D., {Kravtsov}, A.~V., \& {Vikhlinin}, A. 2007, \apj, 668, 1

\bibitem[{{Naiman} {et~al.}(2018){Naiman}, {Pillepich}, {Springel},
  {Ramirez-Ruiz}, {Torrey}, {Vogelsberger}, {Pakmor}, {Nelson}, {Marinacci},
  {Hernquist}, {Weinberger}, \& {Genel}}]{TNG(e)}
{Naiman}, J.~P., {Pillepich}, A., {Springel}, V., {et~al.} 2018, \mnras, 477,
  1206

\bibitem[{{Navarro} {et~al.}(1997){Navarro}, {Frenk}, \& {White}}]{NFW1997}
{Navarro}, J.~F., {Frenk}, C.~S., \& {White}, S. D.~M. 1997, \apj, 490, 493

\bibitem[{{Nelson} {et~al.}(2018){Nelson}, {Pillepich}, {Springel},
  {Weinberger}, {Hernquist}, {Pakmor}, {Genel}, {Torrey}, {Vogelsberger},
  {Kauffmann}, {Marinacci}, \& {Naiman}}]{TNG(d)}
{Nelson}, D., {Pillepich}, A., {Springel}, V., {et~al.} 2018, \mnras, 475, 624

\bibitem[{{Nelson} {et~al.}(2019){Nelson}, {Springel}, {Pillepich},
  {Rodriguez-Gomez}, {Torrey}, {Genel}, {Vogelsberger}, {Pakmor}, {Marinacci},
  {Weinberger}, {Kelley}, {Lovell}, {Diemer}, \& {Hernquist}}]{TNG(a)}
{Nelson}, D., {Springel}, V., {Pillepich}, A., {et~al.} 2019, Computational
  Astrophysics and Cosmology, 6, 2

\bibitem[{{Oukbir} \& {Blanchard}(1992)}]{OukbirBlanchard1992}
{Oukbir}, J. \& {Blanchard}, A. 1992, \aap, 262, L21

\bibitem[{{Pacaud} {et~al.}(2018){Pacaud}, {Pierre}, {Melin}, {Adami},
  {Evrard}, {Galli}, {Gastaldello}, {Maughan}, {Sereno}, {Alis}, {Altieri},
  {Birkinshaw}, {Chiappetti}, {Faccioli}, {Giles}, {Horellou}, {Iovino},
  {Koulouridis}, {Le F{\`e}vre}, {Lidman}, {Lieu}, {Maurogordato},
  {Moscardini}, {Plionis}, {Poggianti}, {Pompei}, {Sadibekova}, {Valtchanov},
  \& {Willis}}]{Pacaud2018-XXLcosmology}
{Pacaud}, F., {Pierre}, M., {Melin}, J.~B., {et~al.} 2018, \aap, 620, A10

\bibitem[{{Pierre} {et~al.}(2016){Pierre}, {Pacaud}, {Adami}, {Alis},
  {Altieri}, {Baran}, {Benoist}, {Birkinshaw}, {Bongiorno}, {Bremer}, {Brusa},
  {Butler}, {Ciliegi}, {Chiappetti}, {Clerc}, {Corasaniti}, {Coupon}, {De
  Breuck}, {Democles}, {Desai}, {Delhaize}, {Devriendt}, {Dubois}, {Eckert},
  {Elyiv}, {Ettori}, {Evrard}, {Faccioli}, {Farahi}, {Ferrari}, {Finet},
  {Fotopoulou}, {Fourmanoit}, {Gandhi}, {Gastaldello}, {Gastaud},
  {Georgantopoulos}, {Giles}, {Guennou}, {Guglielmo}, {Horellou}, {Husband},
  {Huynh}, {Iovino}, {Kilbinger}, {Koulouridis}, {Lavoie}, {Le Brun}, {Le
  Fevre}, {Lidman}, {Lieu}, {Lin}, {Mantz}, {Maughan}, {Maurogordato},
  {McCarthy}, {McGee}, {Melin}, {Melnyk}, {Menanteau}, {Novak}, {Paltani},
  {Plionis}, {Poggianti}, {Pomarede}, {Pompei}, {Ponman}, {Ramos-Ceja},
  {Ranalli}, {Rapetti}, {Raychaudury}, {Reiprich}, {Rottgering}, {Rozo},
  {Rykoff}, {Sadibekova}, {Santos}, {Sauvageot}, {Schimd}, {Sereno}, {Smith},
  {Smol{\v{c}}i{\'c}}, {Snowden}, {Spergel}, {Stanford}, {Surdej}, {Valageas},
  {Valotti}, {Valtchanov}, {Vignali}, {Willis}, \&
  {Ziparo}}]{Pierre2016-XMM-XXL}
{Pierre}, M., {Pacaud}, F., {Adami}, C., {et~al.} 2016, \aap, 592, A1

\bibitem[{{Pillepich} {et~al.}(2018){Pillepich}, {Nelson}, {Hernquist},
  {Springel}, {Pakmor}, {Torrey}, {Weinberger}, {Genel}, {Naiman}, {Marinacci},
  \& {Vogelsberger}}]{TNG(b)}
{Pillepich}, A., {Nelson}, D., {Hernquist}, L., {et~al.} 2018, \mnras, 475, 648

\bibitem[{{Planck Collaboration} {et~al.}(2016{\natexlab{a}}){Planck
  Collaboration}, {Adam}, {Ade}, {Aghanim}, {Arnaud}, {Ashdown}, {Aumont},
  {Baccigalupi}, {Banday}, {Barreiro}, {Bartolo}, {Battaner}, {Benabed},
  {Beno{\^\i}t}, {Benoit-L{\'e}vy}, {Bernard}, {Bersanelli}, {Bertincourt},
  {Bielewicz}, {Bock}, {Bonavera}, {Bond}, {Borrill}, {Bouchet}, {Boulanger},
  {Bucher}, {Burigana}, {Calabrese}, {Cardoso}, {Catalano}, {Challinor},
  {Chamballu}, {Chiang}, {Christensen}, {Clements}, {Colombi}, {Colombo},
  {Combet}, {Couchot}, {Coulais}, {Crill}, {Curto}, {Cuttaia}, {Danese},
  {Davies}, {Davis}, {de Bernardis}, {de Rosa}, {de Zotti}, {Delabrouille},
  {Delouis}, {D{\'e}sert}, {Diego}, {Dole}, {Donzelli}, {Dor{\'e}}, {Douspis},
  {Ducout}, {Dupac}, {Efstathiou}, {Elsner}, {En{\ss}lin}, {Eriksen},
  {Falgarone}, {Fergusson}, {Finelli}, {Forni}, {Frailis}, {Fraisse},
  {Franceschi}, {Frejsel}, {Galeotta}, {Galli}, {Ganga}, {Ghosh}, {Giard},
  {Giraud-H{\'e}raud}, {Gjerl{\o}w}, {Gonz{\'a}lez-Nuevo}, {G{\'o}rski},
  {Gratton}, {Gruppuso}, {Gudmundsson}, {Hansen}, {Hanson}, {Harrison},
  {Henrot-Versill{\'e}}, {Herranz}, {Hildebrandt}, {Hivon}, {Hobson}, {Holmes},
  {Hornstrup}, {Hovest}, {Huffenberger}, {Hurier}, {Jaffe}, {Jaffe}, {Jones},
  {Juvela}, {Keih{\"a}nen}, {Keskitalo}, {Kisner}, {Kneissl}, {Knoche}, {Kunz},
  {Kurki-Suonio}, {Lagache}, {Lamarre}, {Lasenby}, {Lattanzi}, {Lawrence}, {Le
  Jeune}, {Leahy}, {Lellouch}, {Leonardi}, {Lesgourgues}, {Levrier}, {Liguori},
  {Lilje}, {Linden-V{\o}rnle}, {L{\'o}pez-Caniego}, {Lubin},
  {Mac{\'\i}as-P{\'e}rez}, {Maggio}, {Maino}, {Mandolesi}, {Mangilli}, {Maris},
  {Martin}, {Mart{\'\i}nez-Gonz{\'a}lez}, {Masi}, {Matarrese}, {McGehee},
  {Melchiorri}, {Mendes}, {Mennella}, {Migliaccio}, {Mitra},
  {Miville-Desch{\^e}nes}, {Moneti}, {Montier}, {Moreno}, {Morgante},
  {Mortlock}, {Moss}, {Mottet}, {Munshi}, {Murphy}, {Naselsky}, {Nati},
  {Natoli}, {Netterfield}, {N{\o}rgaard-Nielsen}, {Noviello}, {Novikov},
  {Novikov}, {Oxborrow}, {Paci}, {Pagano}, {Pajot}, {Paoletti}, {Pasian},
  {Patanchon}, {Pearson}, {Perdereau}, {Perotto}, {Perrotta}, {Pettorino},
  {Piacentini}, {Piat}, {Pierpaoli}, {Pietrobon}, {Plaszczynski},
  {Pointecouteau}, {Polenta}, {Pratt}, {Pr{\'e}zeau}, {Prunet}, {Puget},
  {Rachen}, {Reinecke}, {Remazeilles}, {Renault}, {Renzi}, {Ristorcelli},
  {Rocha}, {Rosset}, {Rossetti}, {Roudier}, {Rusholme}, {Sandri}, {Santos},
  {Sauv{\'e}}, {Savelainen}, {Savini}, {Scott}, {Seiffert}, {Shellard},
  {Spencer}, {Stolyarov}, {Stompor}, {Sudiwala}, {Sutton}, {Suur-Uski},
  {Sygnet}, {Tauber}, {Terenzi}, {Toffolatti}, {Tomasi}, {Tristram}, {Tucci},
  {Tuovinen}, {Valenziano}, {Valiviita}, {Van Tent}, {Vibert}, {Vielva},
  {Villa}, {Wade}, {Wandelt}, {Watson}, {Wehus}, {Yvon}, {Zacchei}, \&
  {Zonca}}]{PLANCK2016-frequency_maps}
{Planck Collaboration}, {Adam}, R., {Ade}, P.~A.~R., {et~al.}
  2016{\natexlab{a}}, \aap, 594, A8

\bibitem[{{Planck Collaboration} {et~al.}(2014{\natexlab{a}}){Planck
  Collaboration}, {Ade}, {Aghanim}, {Alves}, {Armitage-Caplan}, {Arnaud},
  {Ashdown}, {Atrio-Barandela}, {Aumont}, {Baccigalupi}, {Banday}, {Barreiro},
  {Bartlett}, {Battaner}, {Benabed}, {Beno{\^\i}t}, {Benoit-L{\'e}vy},
  {Bernard}, {Bersanelli}, {Bielewicz}, {Bobin}, {Bock}, {Bonaldi}, {Bond},
  {Borrill}, {Bouchet}, {Boulanger}, {Bridges}, {Bucher}, {Burigana}, {Butler},
  {Cardoso}, {Catalano}, {Chamballu}, {Chary}, {Chen}, {Chiang}, {Chiang},
  {Christensen}, {Church}, {Clements}, {Colombi}, {Colombo}, {Combet},
  {Couchot}, {Coulais}, {Crill}, {Curto}, {Cuttaia}, {Danese}, {Davies}, {de
  Bernardis}, {de Rosa}, {de Zotti}, {Delabrouille}, {Delouis}, {Dempsey},
  {D{\'e}sert}, {Dickinson}, {Diego}, {Dole}, {Donzelli}, {Dor{\'e}},
  {Douspis}, {Dupac}, {Efstathiou}, {En{\ss}lin}, {Eriksen}, {Falgarone},
  {Finelli}, {Forni}, {Frailis}, {Franceschi}, {Fukui}, {Galeotta}, {Ganga},
  {Giard}, {Giraud-H{\'e}raud}, {Gonz{\'a}lez-Nuevo}, {G{\'o}rski}, {Gratton},
  {Gregorio}, {Gruppuso}, {Handa}, {Hansen}, {Hanson}, {Harrison},
  {Henrot-Versill{\'e}}, {Hern{\'a}ndez-Monteagudo}, {Herranz}, {Hildebrandt},
  {Hily-Blant}, {Hivon}, {Hobson}, {Holmes}, {Hornstrup}, {Hovest},
  {Huffenberger}, {Hurier}, {Jaffe}, {Jaffe}, {Jewell}, {Jones}, {Juvela},
  {Keih{\"a}nen}, {Keskitalo}, {Kisner}, {Knoche}, {Knox}, {Kunz},
  {Kurki-Suonio}, {Lagache}, {L{\"a}hteenm{\"a}ki}, {Lamarre}, {Lasenby},
  {Laureijs}, {Lawrence}, {Leonardi}, {Le{\'o}n-Tavares}, {Lesgourgues},
  {Liguori}, {Lilje}, {Linden-V{\o}rnle}, {L{\'o}pez-Caniego}, {Lubin},
  {Mac{\'\i}as-P{\'e}rez}, {Maffei}, {Mandolesi}, {Maris}, {Marshall},
  {Martin}, {Mart{\'\i}nez-Gonz{\'a}lez}, {Masi}, {Massardi}, {Matarrese},
  {Matthai}, {Mazzotta}, {McGehee}, {Melchiorri}, {Mendes}, {Mennella},
  {Migliaccio}, {Mitra}, {Miville-Desch{\^e}nes}, {Moneti}, {Montier}, {Moore},
  {Morgante}, {Morino}, {Mortlock}, {Munshi}, {Murphy}, {Nakajima}, {Naselsky},
  {Nati}, {Natoli}, {Netterfield}, {N{\o}rgaard-Nielsen}, {Noviello},
  {Novikov}, {Novikov}, {Okuda}, {Osborne}, {Oxborrow}, {Paci}, {Pagano},
  {Pajot}, {Paladini}, {Paoletti}, {Pasian}, {Patanchon}, {Perdereau},
  {Perotto}, {Perrotta}, {Piacentini}, {Piat}, {Pierpaoli}, {Pietrobon},
  {Plaszczynski}, {Pointecouteau}, {Polenta}, {Ponthieu}, {Popa}, {Poutanen},
  {Pratt}, {Pr{\'e}zeau}, {Prunet}, {Puget}, {Rachen}, {Reach}, {Rebolo},
  {Reinecke}, {Remazeilles}, {Renault}, {Ricciardi}, {Riller}, {Ristorcelli},
  {Rocha}, {Rosset}, {Roudier}, {Rowan-Robinson}, {Rubi{\~n}o-Mart{\'\i}n},
  {Rusholme}, {Sandri}, {Santos}, {Savini}, {Scott}, {Seiffert}, {Shellard},
  {Spencer}, {Starck}, {Stolyarov}, {Stompor}, {Sudiwala}, {Sunyaev}, {Sureau},
  {Sutton}, {Suur-Uski}, {Sygnet}, {Tauber}, {Tavagnacco}, {Terenzi}, {Thomas},
  {Toffolatti}, {Tomasi}, {Torii}, {Tristram}, {Tucci}, {Tuovinen}, {Umana},
  {Valenziano}, {Valiviita}, {Van Tent}, {Vielva}, {Villa}, {Vittorio}, {Wade},
  {Wandelt}, {Wehus}, {Yamamoto}, {Yoda}, {Yvon}, {Zacchei}, \&
  {Zonca}}]{PLANCK2014-CO_emission}
{Planck Collaboration}, {Ade}, P.~A.~R., {Aghanim}, N., {et~al.}
  2014{\natexlab{a}}, \aap, 571, A13

\bibitem[{{Planck Collaboration} {et~al.}(2016{\natexlab{b}}){Planck
  Collaboration}, {Ade}, {Aghanim}, {Arg{\"u}eso}, {Arnaud}, {Ashdown},
  {Aumont}, {Baccigalupi}, {Banday}, {Barreiro}, {Bartolo}, {Battaner},
  {Beichman}, {Benabed}, {Beno{\^\i}t}, {Benoit-L{\'e}vy}, {Bernard},
  {Bersanelli}, {Bielewicz}, {Bock}, {B{\"o}hringer}, {Bonaldi}, {Bonavera},
  {Bond}, {Borrill}, {Bouchet}, {Boulanger}, {Bucher}, {Burigana}, {Butler},
  {Calabrese}, {Cardoso}, {Carvalho}, {Catalano}, {Challinor}, {Chamballu},
  {Chary}, {Chiang}, {Christensen}, {Clemens}, {Clements}, {Colombi},
  {Colombo}, {Combet}, {Couchot}, {Coulais}, {Crill}, {Curto}, {Cuttaia},
  {Danese}, {Davies}, {Davis}, {de Bernardis}, {de Rosa}, {de Zotti},
  {Delabrouille}, {D{\'e}sert}, {Dickinson}, {Diego}, {Dole}, {Donzelli},
  {Dor{\'e}}, {Douspis}, {Ducout}, {Dupac}, {Efstathiou}, {Elsner},
  {En{\ss}lin}, {Eriksen}, {Falgarone}, {Fergusson}, {Finelli}, {Forni},
  {Frailis}, {Fraisse}, {Franceschi}, {Frejsel}, {Galeotta}, {Galli}, {Ganga},
  {Giard}, {Giraud-H{\'e}raud}, {Gjerl{\o}w}, {Gonz{\'a}lez-Nuevo},
  {G{\'o}rski}, {Gratton}, {Gregorio}, {Gruppuso}, {Gudmundsson}, {Hansen},
  {Hanson}, {Harrison}, {Helou}, {Henrot-Versill{\'e}},
  {Hern{\'a}ndez-Monteagudo}, {Herranz}, {Hildebrandt}, {Hivon}, {Hobson},
  {Holmes}, {Hornstrup}, {Hovest}, {Huffenberger}, {Hurier}, {Jaffe}, {Jaffe},
  {Jones}, {Juvela}, {Keih{\"a}nen}, {Keskitalo}, {Kisner}, {Kneissl},
  {Knoche}, {Kunz}, {Kurki-Suonio}, {Lagache}, {L{\"a}hteenm{\"a}ki},
  {Lamarre}, {Lasenby}, {Lattanzi}, {Lawrence}, {Leahy}, {Leonardi},
  {Le{\'o}n-Tavares}, {Lesgourgues}, {Levrier}, {Liguori}, {Lilje},
  {Linden-V{\o}rnle}, {L{\'o}pez-Caniego}, {Lubin}, {Mac{\'\i}as-P{\'e}rez},
  {Maggio}, {Maino}, {Mandolesi}, {Mangilli}, {Maris}, {Marshall}, {Martin},
  {Mart{\'\i}nez-Gonz{\'a}lez}, {Masi}, {Matarrese}, {McGehee}, {Meinhold},
  {Melchiorri}, {Mendes}, {Mennella}, {Migliaccio}, {Mitra},
  {Miville-Desch{\^e}nes}, {Moneti}, {Montier}, {Morgante}, {Mortlock}, {Moss},
  {Munshi}, {Murphy}, {Naselsky}, {Nati}, {Natoli}, {Negrello}, {Netterfield},
  {N{\o}rgaard-Nielsen}, {Noviello}, {Novikov}, {Novikov}, {Oxborrow}, {Paci},
  {Pagano}, {Pajot}, {Paladini}, {Paoletti}, {Partridge}, {Pasian},
  {Patanchon}, {Pearson}, {Perdereau}, {Perotto}, {Perrotta}, {Pettorino},
  {Piacentini}, {Piat}, {Pierpaoli}, {Pietrobon}, {Plaszczynski},
  {Pointecouteau}, {Polenta}, {Pratt}, {Pr{\'e}zeau}, {Prunet}, {Puget},
  {Rachen}, {Reach}, {Rebolo}, {Reinecke}, {Remazeilles}, {Renault}, {Renzi},
  {Ristorcelli}, {Rocha}, {Rosset}, {Rossetti}, {Roudier}, {Rowan-Robinson},
  {Rubi{\~n}o-Mart{\'\i}n}, {Rusholme}, {Sandri}, {Sanghera}, {Santos},
  {Savelainen}, {Savini}, {Scott}, {Seiffert}, {Shellard}, {Spencer},
  {Stolyarov}, {Sudiwala}, {Sunyaev}, {Sutton}, {Suur-Uski}, {Sygnet},
  {Tauber}, {Terenzi}, {Toffolatti}, {Tomasi}, {Tornikoski}, {Tristram},
  {Tucci}, {Tuovinen}, {T{\"u}rler}, {Umana}, {Valenziano}, {Valiviita}, {Van
  Tent}, {Vielva}, {Villa}, {Wade}, {Walter}, {Wandelt}, {Wehus}, {Yvon},
  {Zacchei}, \& {Zonca}}]{PLANCK2016-point_sources_catalog2}
{Planck Collaboration}, {Ade}, P.~A.~R., {Aghanim}, N., {et~al.}
  2016{\natexlab{b}}, \aap, 594, A26

\bibitem[{{Planck Collaboration} {et~al.}(2014{\natexlab{b}}){Planck
  Collaboration}, {Ade}, {Aghanim}, {Armitage-Caplan}, {Arnaud}, {Ashdown},
  {Atrio-Barandela}, {Aumont}, {Aussel}, {Baccigalupi}, {Banday}, {Barreiro},
  {Barrena}, {Bartelmann}, {Bartlett}, {Battaner}, {Benabed}, {Beno{\^\i}t},
  {Benoit-L{\'e}vy}, {Bernard}, {Bersanelli}, {Bielewicz}, {Bikmaev}, {Bobin},
  {Bock}, {B{\"o}hringer}, {Bonaldi}, {Bond}, {Borrill}, {Bouchet}, {Bridges},
  {Bucher}, {Burenin}, {Burigana}, {Butler}, {Cardoso}, {Carvalho}, {Catalano},
  {Challinor}, {Chamballu}, {Chary}, {Chen}, {Chiang}, {Chiang}, {Chon},
  {Christensen}, {Churazov}, {Church}, {Clements}, {Colombi}, {Colombo},
  {Comis}, {Couchot}, {Coulais}, {Crill}, {Curto}, {Cuttaia}, {Da Silva},
  {Dahle}, {Danese}, {Davies}, {Davis}, {de Bernardis}, {de Rosa}, {de Zotti},
  {Delabrouille}, {Delouis}, {D{\'e}mocl{\`e}s}, {D{\'e}sert}, {Dickinson},
  {Diego}, {Dolag}, {Dole}, {Donzelli}, {Dor{\'e}}, {Douspis}, {Dupac},
  {Efstathiou}, {Eisenhardt}, {En{\ss}lin}, {Eriksen}, {Feroz}, {Finelli},
  {Flores-Cacho}, {Forni}, {Frailis}, {Franceschi}, {Fromenteau}, {Galeotta},
  {Ganga}, {G{\'e}nova-Santos}, {Giard}, {Giardino}, {Gilfanov},
  {Giraud-H{\'e}raud}, {Gonz{\'a}lez-Nuevo}, {G{\'o}rski}, {Grainge},
  {Gratton}, {Gregorio}, {Groeneboom}, {Gruppuso}, {Hansen}, {Hanson},
  {Harrison}, {Hempel}, {Henrot-Versill{\'e}}, {Hern{\'a}ndez-Monteagudo},
  {Herranz}, {Hildebrandt}, {Hivon}, {Hobson}, {Holmes}, {Hornstrup}, {Hovest},
  {Huffenberger}, {Hurier}, {Hurley-Walker}, {Jaffe}, {Jaffe}, {Jones},
  {Juvela}, {Keih{\"a}nen}, {Keskitalo}, {Khamitov}, {Kisner}, {Kneissl},
  {Knoche}, {Knox}, {Kunz}, {Kurki-Suonio}, {Lagache}, {L{\"a}hteenm{\"a}ki},
  {Lamarre}, {Lasenby}, {Laureijs}, {Lawrence}, {Leahy}, {Leonardi},
  {Le{\'o}n-Tavares}, {Lesgourgues}, {Li}, {Liddle}, {Liguori}, {Lilje},
  {Linden-V{\o}rnle}, {L{\'o}pez-Caniego}, {Lubin}, {Mac{\'\i}as-P{\'e}rez},
  {MacTavish}, {Maffei}, {Maino}, {Mandolesi}, {Maris}, {Marshall}, {Martin},
  {Mart{\'\i}nez-Gonz{\'a}lez}, {Masi}, {Massardi}, {Matarrese}, {Matthai},
  {Mazzotta}, {Mei}, {Meinhold}, {Melchiorri}, {Melin}, {Mendes}, {Mennella},
  {Migliaccio}, {Mikkelsen}, {Mitra}, {Miville-Desch{\^e}nes}, {Moneti},
  {Montier}, {Morgante}, {Mortlock}, {Munshi}, {Murphy}, {Naselsky}, {Nati},
  {Natoli}, {Nesvadba}, {Netterfield}, {N{\o}rgaard-Nielsen}, {Noviello},
  {Novikov}, {Novikov}, {O'Dwyer}, {Olamaie}, {Osborne}, {Oxborrow}, {Paci},
  {Pagano}, {Pajot}, {Paoletti}, {Pasian}, {Patanchon}, {Pearson}, {Perdereau},
  {Perotto}, {Perrott}, {Perrotta}, {Piacentini}, {Piat}, {Pierpaoli},
  {Pietrobon}, {Plaszczynski}, {Pointecouteau}, {Polenta}, {Ponthieu}, {Popa},
  {Poutanen}, {Pratt}, {Pr{\'e}zeau}, {Prunet}, {Puget}, {Rachen}, {Reach},
  {Rebolo}, {Reinecke}, {Remazeilles}, {Renault}, {Ricciardi}, {Riller},
  {Ristorcelli}, {Rocha}, {Rosset}, {Roudier}, {Rowan-Robinson},
  {Rubi{\~n}o-Mart{\'\i}n}, {Rumsey}, {Rusholme}, {Sandri}, {Santos},
  {Saunders}, {Savini}, {Schammel}, {Scott}, {Seiffert}, {Shellard},
  {Shimwell}, {Spencer}, {Stanford}, {Starck}, {Stolyarov}, {Stompor},
  {Sudiwala}, {Sunyaev}, {Sureau}, {Sutton}, {Suur-Uski}, {Sygnet}, {Tauber},
  {Tavagnacco}, {Terenzi}, {Toffolatti}, {Tomasi}, {Tristram}, {Tucci},
  {Tuovinen}, {T{\"u}rler}, {Umana}, {Valenziano}, {Valiviita}, {Van Tent},
  {Vibert}, {Vielva}, {Villa}, {Vittorio}, {Wade}, {Wandelt}, {White}, {White},
  {Yvon}, {Zacchei}, \& {Zonca}}]{PLANCK2014-PSZ}
{Planck Collaboration}, {Ade}, P.~A.~R., {Aghanim}, N., {et~al.}
  2014{\natexlab{b}}, \aap, 571, A29

\bibitem[{{Planck Collaboration} {et~al.}(2014{\natexlab{c}}){Planck
  Collaboration}, {Ade}, {Aghanim}, {Armitage-Caplan}, {Arnaud}, {Ashdown},
  {Atrio-Barandela}, {Aumont}, {Baccigalupi}, {Banday}, {Barreiro}, {Barrena},
  {Bartlett}, {Battaner}, {Battye}, {Benabed}, {Beno{\^\i}t},
  {Benoit-L{\'e}vy}, {Bernard}, {Bersanelli}, {Bielewicz}, {Bikmaev},
  {Blanchard}, {Bobin}, {Bock}, {B{\"o}hringer}, {Bonaldi}, {Bond}, {Borrill},
  {Bouchet}, {Bourdin}, {Bridges}, {Brown}, {Bucher}, {Burenin}, {Burigana},
  {Butler}, {Cardoso}, {Carvalho}, {Catalano}, {Challinor}, {Chamballu},
  {Chary}, {Chiang}, {Chiang}, {Chon}, {Christensen}, {Church}, {Clements},
  {Colombi}, {Colombo}, {Couchot}, {Coulais}, {Crill}, {Curto}, {Cuttaia}, {Da
  Silva}, {Dahle}, {Danese}, {Davies}, {Davis}, {de Bernardis}, {de Rosa}, {de
  Zotti}, {Delabrouille}, {Delouis}, {D{\'e}mocl{\`e}s}, {D{\'e}sert},
  {Dickinson}, {Diego}, {Dolag}, {Dole}, {Donzelli}, {Dor{\'e}}, {Douspis},
  {Dupac}, {Efstathiou}, {En{\ss}lin}, {Eriksen}, {Finelli}, {Flores-Cacho},
  {Forni}, {Frailis}, {Franceschi}, {Fromenteau}, {Galeotta}, {Ganga},
  {G{\'e}nova-Santos}, {Giard}, {Giardino}, {Giraud-H{\'e}raud},
  {Gonz{\'a}lez-Nuevo}, {G{\'o}rski}, {Gratton}, {Gregorio}, {Gruppuso},
  {Hansen}, {Hanson}, {Harrison}, {Henrot-Versill{\'e}},
  {Hern{\'a}ndez-Monteagudo}, {Herranz}, {Hildebrandt}, {Hivon}, {Hobson},
  {Holmes}, {Hornstrup}, {Hovest}, {Huffenberger}, {Hurier}, {Jaffe}, {Jaffe},
  {Jones}, {Juvela}, {Keih{\"a}nen}, {Keskitalo}, {Khamitov}, {Kisner},
  {Kneissl}, {Knoche}, {Knox}, {Kunz}, {Kurki-Suonio}, {Lagache},
  {L{\"a}hteenm{\"a}ki}, {Lamarre}, {Lasenby}, {Laureijs}, {Lawrence}, {Leahy},
  {Leonardi}, {Le{\'o}n-Tavares}, {Lesgourgues}, {Liddle}, {Liguori}, {Lilje},
  {Linden-V{\o}rnle}, {L{\'o}pez-Caniego}, {Lubin}, {Mac{\'\i}as-P{\'e}rez},
  {Maffei}, {Maino}, {Mandolesi}, {Marcos-Caballero}, {Maris}, {Marshall},
  {Martin}, {Mart{\'\i}nez-Gonz{\'a}lez}, {Masi}, {Matarrese}, {Matthai},
  {Mazzotta}, {Meinhold}, {Melchiorri}, {Melin}, {Mendes}, {Mennella},
  {Migliaccio}, {Mitra}, {Miville-Desch{\^e}nes}, {Moneti}, {Montier},
  {Morgante}, {Mortlock}, {Moss}, {Munshi}, {Naselsky}, {Nati}, {Natoli},
  {Netterfield}, {N{\o}rgaard-Nielsen}, {Noviello}, {Novikov}, {Novikov},
  {Osborne}, {Oxborrow}, {Paci}, {Pagano}, {Pajot}, {Paoletti}, {Partridge},
  {Pasian}, {Patanchon}, {Perdereau}, {Perotto}, {Perrotta}, {Piacentini},
  {Piat}, {Pierpaoli}, {Pietrobon}, {Plaszczynski}, {Pointecouteau}, {Polenta},
  {Ponthieu}, {Popa}, {Poutanen}, {Pratt}, {Pr{\'e}zeau}, {Prunet}, {Puget},
  {Rachen}, {Rebolo}, {Reinecke}, {Remazeilles}, {Renault}, {Ricciardi},
  {Riller}, {Ristorcelli}, {Rocha}, {Roman}, {Rosset}, {Roudier},
  {Rowan-Robinson}, {Rubi{\~n}o-Mart{\'\i}n}, {Rusholme}, {Sandri}, {Santos},
  {Savini}, {Scott}, {Seiffert}, {Shellard}, {Spencer}, {Starck}, {Stolyarov},
  {Stompor}, {Sudiwala}, {Sunyaev}, {Sureau}, {Sutton}, {Suur-Uski}, {Sygnet},
  {Tauber}, {Tavagnacco}, {Terenzi}, {Toffolatti}, {Tomasi}, {Tristram},
  {Tucci}, {Tuovinen}, {T{\"u}rler}, {Umana}, {Valenziano}, {Valiviita}, {Van
  Tent}, {Vielva}, {Villa}, {Vittorio}, {Wade}, {Wandelt}, {Weller}, {White},
  {White}, {Yvon}, {Zacchei}, \& {Zonca}}]{PLANCK2014-PSZ_cosmology}
{Planck Collaboration}, {Ade}, P.~A.~R., {Aghanim}, N., {et~al.}
  2014{\natexlab{c}}, \aap, 571, A20

\bibitem[{{Planck Collaboration} {et~al.}(2014{\natexlab{d}}){Planck
  Collaboration}, {Ade}, {Aghanim}, {Armitage-Caplan}, {Arnaud}, {Ashdown},
  {Atrio-Barandela}, {Aumont}, {Baccigalupi}, {Banday}, {Barreiro}, {Bartlett},
  {Battaner}, {Benabed}, {Beno{\^\i}t}, {Benoit-L{\'e}vy}, {Bernard},
  {Bersanelli}, {Bielewicz}, {Bobin}, {Bock}, {Bonaldi}, {Bond}, {Borrill},
  {Bouchet}, {Bridges}, {Bucher}, {Burigana}, {Butler}, {Calabrese},
  {Cappellini}, {Cardoso}, {Catalano}, {Challinor}, {Chamballu}, {Chary},
  {Chen}, {Chiang}, {Chiang}, {Christensen}, {Church}, {Clements}, {Colombi},
  {Colombo}, {Couchot}, {Coulais}, {Crill}, {Curto}, {Cuttaia}, {Danese},
  {Davies}, {Davis}, {de Bernardis}, {de Rosa}, {de Zotti}, {Delabrouille},
  {Delouis}, {D{\'e}sert}, {Dickinson}, {Diego}, {Dolag}, {Dole}, {Donzelli},
  {Dor{\'e}}, {Douspis}, {Dunkley}, {Dupac}, {Efstathiou}, {Elsner},
  {En{\ss}lin}, {Eriksen}, {Finelli}, {Forni}, {Frailis}, {Fraisse},
  {Franceschi}, {Gaier}, {Galeotta}, {Galli}, {Ganga}, {Giard}, {Giardino},
  {Giraud-H{\'e}raud}, {Gjerl{\o}w}, {Gonz{\'a}lez-Nuevo}, {G{\'o}rski},
  {Gratton}, {Gregorio}, {Gruppuso}, {Gudmundsson}, {Haissinski}, {Hamann},
  {Hansen}, {Hanson}, {Harrison}, {Henrot-Versill{\'e}},
  {Hern{\'a}ndez-Monteagudo}, {Herranz}, {Hildebrandt}, {Hivon}, {Hobson},
  {Holmes}, {Hornstrup}, {Hou}, {Hovest}, {Huffenberger}, {Jaffe}, {Jaffe},
  {Jewell}, {Jones}, {Juvela}, {Keih{\"a}nen}, {Keskitalo}, {Kisner},
  {Kneissl}, {Knoche}, {Knox}, {Kunz}, {Kurki-Suonio}, {Lagache},
  {L{\"a}hteenm{\"a}ki}, {Lamarre}, {Lasenby}, {Lattanzi}, {Laureijs},
  {Lawrence}, {Leach}, {Leahy}, {Leonardi}, {Le{\'o}n-Tavares}, {Lesgourgues},
  {Lewis}, {Liguori}, {Lilje}, {Linden-V{\o}rnle}, {L{\'o}pez-Caniego},
  {Lubin}, {Mac{\'\i}as-P{\'e}rez}, {Maffei}, {Maino}, {Mandolesi}, {Maris},
  {Marshall}, {Martin}, {Mart{\'\i}nez-Gonz{\'a}lez}, {Masi}, {Massardi},
  {Matarrese}, {Matthai}, {Mazzotta}, {Meinhold}, {Melchiorri}, {Melin},
  {Mendes}, {Menegoni}, {Mennella}, {Migliaccio}, {Millea}, {Mitra},
  {Miville-Desch{\^e}nes}, {Moneti}, {Montier}, {Morgante}, {Mortlock}, {Moss},
  {Munshi}, {Murphy}, {Naselsky}, {Nati}, {Natoli}, {Netterfield},
  {N{\o}rgaard-Nielsen}, {Noviello}, {Novikov}, {Novikov}, {O'Dwyer},
  {Osborne}, {Oxborrow}, {Paci}, {Pagano}, {Pajot}, {Paladini}, {Paoletti},
  {Partridge}, {Pasian}, {Patanchon}, {Pearson}, {Pearson}, {Peiris},
  {Perdereau}, {Perotto}, {Perrotta}, {Pettorino}, {Piacentini}, {Piat},
  {Pierpaoli}, {Pietrobon}, {Plaszczynski}, {Platania}, {Pointecouteau},
  {Polenta}, {Ponthieu}, {Popa}, {Poutanen}, {Pratt}, {Pr{\'e}zeau}, {Prunet},
  {Puget}, {Rachen}, {Reach}, {Rebolo}, {Reinecke}, {Remazeilles}, {Renault},
  {Ricciardi}, {Riller}, {Ristorcelli}, {Rocha}, {Rosset}, {Roudier},
  {Rowan-Robinson}, {Rubi{\~n}o-Mart{\'\i}n}, {Rusholme}, {Sandri}, {Santos},
  {Savelainen}, {Savini}, {Scott}, {Seiffert}, {Shellard}, {Spencer}, {Starck},
  {Stolyarov}, {Stompor}, {Sudiwala}, {Sunyaev}, {Sureau}, {Sutton},
  {Suur-Uski}, {Sygnet}, {Tauber}, {Tavagnacco}, {Terenzi}, {Toffolatti},
  {Tomasi}, {Tristram}, {Tucci}, {Tuovinen}, {T{\"u}rler}, {Umana},
  {Valenziano}, {Valiviita}, {Van Tent}, {Vielva}, {Villa}, {Vittorio}, {Wade},
  {Wandelt}, {Wehus}, {White}, {White}, {Wilkinson}, {Yvon}, {Zacchei}, \&
  {Zonca}}]{PLANCK2014-Cosmology}
{Planck Collaboration}, {Ade}, P.~A.~R., {Aghanim}, N., {et~al.}
  2014{\natexlab{d}}, \aap, 571, A16

\bibitem[{{Planck Collaboration} {et~al.}(2014{\natexlab{e}}){Planck
  Collaboration}, {Ade}, {Aghanim}, {Armitage-Caplan}, {Arnaud}, {Ashdown},
  {Atrio-Barandela}, {Aumont}, {Baccigalupi}, {Banday}, {Barreiro}, {Battaner},
  {Benabed}, {Beno{\^\i}t}, {Benoit-L{\'e}vy}, {Bernard}, {Bersanelli},
  {Bielewicz}, {Bobin}, {Bock}, {Bond}, {Borrill}, {Bouchet}, {Bowyer},
  {Bridges}, {Bucher}, {Burigana}, {Cardoso}, {Catalano}, {Challinor},
  {Chamballu}, {Chary}, {Chiang}, {Chiang}, {Christensen}, {Church},
  {Clements}, {Colombi}, {Colombo}, {Couchot}, {Coulais}, {Crill}, {Curto},
  {Cuttaia}, {Danese}, {Davies}, {de Bernardis}, {de Rosa}, {de Zotti},
  {Delabrouille}, {Delouis}, {D{\'e}sert}, {Diego}, {Dole}, {Donzelli},
  {Dor{\'e}}, {Douspis}, {Dunkley}, {Dupac}, {Efstathiou}, {En{\ss}lin},
  {Eriksen}, {Finelli}, {Forni}, {Frailis}, {Fraisse}, {Franceschi},
  {Galeotta}, {Ganga}, {Giard}, {Giraud-H{\'e}raud}, {Gonz{\'a}lez-Nuevo},
  {G{\'o}rski}, {Gratton}, {Gregorio}, {Gruppuso}, {Gudmundsson}, {Haissinski},
  {Hansen}, {Hanson}, {Harrison}, {Henrot-Versill{\'e}},
  {Hern{\'a}ndez-Monteagudo}, {Herranz}, {Hildebrandt}, {Hivon}, {Hobson},
  {Holmes}, {Hornstrup}, {Hou}, {Hovest}, {Huffenberger}, {Jaffe}, {Jaffe},
  {Jones}, {Juvela}, {Keih{\"a}nen}, {Keskitalo}, {Kisner}, {Kneissl},
  {Knoche}, {Knox}, {Kunz}, {Kurki-Suonio}, {Lagache}, {Lamarre}, {Lasenby},
  {Laureijs}, {Lawrence}, {Leonardi}, {Leroy}, {Lesgourgues}, {Liguori},
  {Lilje}, {Linden-V{\o}rnle}, {L{\'o}pez-Caniego}, {Lubin},
  {Mac{\'\i}as-P{\'e}rez}, {MacTavish}, {Maffei}, {Mandolesi}, {Maris},
  {Marshall}, {Martin}, {Mart{\'\i}nez-Gonz{\'a}lez}, {Masi}, {Massardi},
  {Matarrese}, {Matsumura}, {Matthai}, {Mazzotta}, {McGehee}, {Melchiorri},
  {Mendes}, {Mennella}, {Migliaccio}, {Mitra}, {Miville-Desch{\^e}nes},
  {Moneti}, {Montier}, {Morgante}, {Mortlock}, {Munshi}, {Murphy}, {Naselsky},
  {Nati}, {Natoli}, {Netterfield}, {N{\o}rgaard-Nielsen}, {Noviello},
  {Novikov}, {Novikov}, {Osborne}, {Oxborrow}, {Paci}, {Pagano}, {Pajot},
  {Paoletti}, {Pasian}, {Patanchon}, {Perdereau}, {Perotto}, {Perrotta},
  {Piacentini}, {Piat}, {Pierpaoli}, {Pietrobon}, {Plaszczynski},
  {Pointecouteau}, {Polegre}, {Polenta}, {Ponthieu}, {Popa}, {Poutanen},
  {Pratt}, {Pr{\'e}zeau}, {Prunet}, {Puget}, {Rachen}, {Reinecke},
  {Remazeilles}, {Renault}, {Ricciardi}, {Riller}, {Ristorcelli}, {Rocha},
  {Rosset}, {Roudier}, {Rowan-Robinson}, {Rusholme}, {Sandri}, {Santos},
  {Sauv{\'e}}, {Savini}, {Scott}, {Shellard}, {Spencer}, {Starck}, {Stolyarov},
  {Stompor}, {Sudiwala}, {Sureau}, {Sutton}, {Suur-Uski}, {Sygnet}, {Tauber},
  {Tavagnacco}, {Terenzi}, {Tomasi}, {Tristram}, {Tucci}, {Umana},
  {Valenziano}, {Valiviita}, {Van Tent}, {Vielva}, {Villa}, {Vittorio}, {Wade},
  {Wandelt}, {Yvon}, {Zacchei}, \& {Zonca}}]{Planck2014-HFI_beams}
{Planck Collaboration}, {Ade}, P.~A.~R., {Aghanim}, N., {et~al.}
  2014{\natexlab{e}}, \aap, 571, A7

\bibitem[{{Planck Collaboration} {et~al.}(2013){Planck Collaboration}, {Ade},
  {Aghanim}, {Arnaud}, {Ashdown}, {Atrio-Barandela}, {Aumont}, {Baccigalupi},
  {Balbi}, {Banday}, {Barreiro}, {Bartlett}, {Battaner}, {Benabed},
  {Beno{\^\i}t}, {Bernard}, {Bersanelli}, {Bhatia}, {Bikmaev}, {Bobin},
  {B{\"o}hringer}, {Bonaldi}, {Bond}, {Borgani}, {Borrill}, {Bouchet},
  {Bourdin}, {Brown}, {Burenin}, {Burigana}, {Cabella}, {Cardoso}, {Carvalho},
  {Castex}, {Catalano}, {Cay{\'o}n}, {Chamballu}, {Chiang}, {Chon},
  {Christensen}, {Churazov}, {Clements}, {Colafrancesco}, {Colombi}, {Colombo},
  {Comis}, {Coulais}, {Crill}, {Cuttaia}, {Da Silva}, {Dahle}, {Danese},
  {Davis}, {de Bernardis}, {de Gasperis}, {de Zotti}, {Delabrouille},
  {D{\'e}mocl{\`e}s}, {D{\'e}sert}, {Diego}, {Dolag}, {Dole}, {Donzelli},
  {Dor{\'e}}, {D{\"o}rl}, {Douspis}, {Dupac}, {Efstathiou}, {En{\ss}lin},
  {Eriksen}, {Finelli}, {Flores-Cacho}, {Forni}, {Fosalba}, {Frailis},
  {Franceschi}, {Frommert}, {Galeotta}, {Ganga}, {G{\'e}nova-Santos}, {Giard},
  {Giraud-H{\'e}raud}, {Gonz{\'a}lez-Nuevo}, {G{\'o}rski}, {Gregorio},
  {Gruppuso}, {Hansen}, {Harrison}, {Hempel}, {Henrot-Versill{\'e}},
  {Hern{\'a}ndez-Monteagudo}, {Herranz}, {Hildebrandt}, {Hivon}, {Hobson},
  {Holmes}, {Hurier}, {Jaffe}, {Jaffe}, {Jagemann}, {Jones}, {Juvela},
  {Keih{\"a}nen}, {Khamitov}, {Kisner}, {Kneissl}, {Knoche}, {Knox}, {Kunz},
  {Kurki-Suonio}, {Lagache}, {L{\"a}hteenm{\"a}ki}, {Lamarre}, {Lasenby},
  {Lawrence}, {Le Jeune}, {Leonardi}, {Liddle}, {Lilje}, {L{\'o}pez-Caniego},
  {Luzzi}, {Mac{\'\i}as-P{\'e}rez}, {Maino}, {Mandolesi}, {Maris}, {Marleau},
  {Marshall}, {Mart{\'\i}nez-Gonz{\'a}lez}, {Masi}, {Massardi}, {Matarrese},
  {Mazzotta}, {Mei}, {Melchiorri}, {Melin}, {Mendes}, {Mennella}, {Mitra},
  {Miville-Desch{\^e}nes}, {Moneti}, {Montier}, {Morgante}, {Mortlock},
  {Munshi}, {Murphy}, {Naselsky}, {Nati}, {Natoli}, {N{\o}rgaard-Nielsen},
  {Noviello}, {Novikov}, {Novikov}, {Osborne}, {Pajot}, {Paoletti}, {Pasian},
  {Patanchon}, {Perdereau}, {Perotto}, {Perrotta}, {Piacentini}, {Piat},
  {Pierpaoli}, {Piffaretti}, {Plaszczynski}, {Pointecouteau}, {Polenta},
  {Ponthieu}, {Popa}, {Poutanen}, {Pratt}, {Prunet}, {Puget}, {Rachen},
  {Reach}, {Rebolo}, {Reinecke}, {Remazeilles}, {Renault}, {Ricciardi},
  {Riller}, {Ristorcelli}, {Rocha}, {Roman}, {Rosset}, {Rossetti},
  {Rubi{\~n}o-Mart{\'\i}n}, {Rusholme}, {Sandri}, {Savini}, {Scott}, {Smoot},
  {Starck}, {Sudiwala}, {Sunyaev}, {Sutton}, {Suur-Uski}, {Sygnet}, {Tauber},
  {Terenzi}, {Toffolatti}, {Tomasi}, {Tristram}, {Tuovinen}, {Valenziano}, {Van
  Tent}, {Varis}, {Vielva}, {Villa}, {Vittorio}, {Wade}, {Wandelt}, {Welikala},
  {White}, {White}, {Yvon}, {Zacchei}, \& {Zonca}}]{PLANCK2013-profile}
{Planck Collaboration}, {Ade}, P.~A.~R., {Aghanim}, N., {et~al.} 2013, \aap,
  550, A131

\bibitem[{{Planck Collaboration} {et~al.}(2011){Planck Collaboration}, {Ade},
  {Aghanim}, {Arnaud}, {Ashdown}, {Aumont}, {Baccigalupi}, {Balbi}, {Banday},
  {Barreiro}, {Bartelmann}, {Bartlett}, {Battaner}, {Battye}, {Benabed},
  {Beno{\^\i}t}, {Bernard}, {Bersanelli}, {Bhatia}, {Bock}, {Bonaldi}, {Bond},
  {Borrill}, {Bouchet}, {Brown}, {Bucher}, {Burigana}, {Cabella}, {Cantalupo},
  {Cardoso}, {Carvalho}, {Catalano}, {Cay{\'o}n}, {Challinor}, {Chamballu},
  {Chary}, {Chiang}, {Chiang}, {Chon}, {Christensen}, {Churazov}, {Clements},
  {Colafrancesco}, {Colombi}, {Couchot}, {Coulais}, {Crill}, {Cuttaia}, {da
  Silva}, {Dahle}, {Danese}, {Davis}, {de Bernardis}, {de Gasperis}, {de Rosa},
  {de Zotti}, {Delabrouille}, {Delouis}, {D{\'e}sert}, {Dickinson}, {Diego},
  {Dolag}, {Dole}, {Donzelli}, {Dor{\'e}}, {D{\"o}rl}, {Douspis}, {Dupac},
  {Efstathiou}, {Eisenhardt}, {En{\ss}lin}, {Feroz}, {Finelli}, {Flores-Cacho},
  {Forni}, {Fosalba}, {Frailis}, {Franceschi}, {Fromenteau}, {Galeotta},
  {Ganga}, {G{\'e}nova-Santos}, {Giard}, {Giardino}, {Giraud-H{\'e}raud},
  {Gonz{\'a}lez-Nuevo}, {Gonz{\'a}lez-Riestra}, {G{\'o}rski}, {Grainge},
  {Gratton}, {Gregorio}, {Gruppuso}, {Harrison}, {Hein{\"a}m{\"a}ki},
  {Henrot-Versill{\'e}}, {Hern{\'a}ndez-Monteagudo}, {Herranz}, {Hildebrandt},
  {Hivon}, {Hobson}, {Holmes}, {Hovest}, {Hoyland}, {Huffenberger}, {Hurier},
  {Hurley-Walker}, {Jaffe}, {Jones}, {Juvela}, {Keih{\"a}nen}, {Keskitalo},
  {Kisner}, {Kneissl}, {Knox}, {Kurki-Suonio}, {Lagache}, {Lamarre}, {Lasenby},
  {Laureijs}, {Lawrence}, {Le Jeune}, {Leach}, {Leonardi}, {Li}, {Liddle},
  {Lilje}, {Linden-V{\o}rnle}, {L{\'o}pez-Caniego}, {Lubin},
  {Mac{\'\i}as-P{\'e}rez}, {MacTavish}, {Maffei}, {Maino}, {Mandolesi}, {Mann},
  {Maris}, {Marleau}, {Mart{\'\i}nez-Gonz{\'a}lez}, {Masi}, {Matarrese},
  {Matthai}, {Mazzotta}, {Mei}, {Meinhold}, {Melchiorri}, {Melin}, {Mendes},
  {Mennella}, {Mitra}, {Miville-Desch{\^e}nes}, {Moneti}, {Montier},
  {Morgante}, {Mortlock}, {Munshi}, {Murphy}, {Naselsky}, {Nati}, {Natoli},
  {Netterfield}, {N{\o}rgaard-Nielsen}, {Noviello}, {Novikov}, {Novikov},
  {Olamaie}, {Osborne}, {Pajot}, {Pasian}, {Patanchon}, {Pearson}, {Perdereau},
  {Perotto}, {Perrotta}, {Piacentini}, {Piat}, {Pierpaoli}, {Piffaretti},
  {Plaszczynski}, {Pointecouteau}, {Polenta}, {Ponthieu}, {Poutanen}, {Pratt},
  {Pr{\'e}zeau}, {Prunet}, {Puget}, {Rachen}, {Reach}, {Rebolo}, {Reinecke},
  {Renault}, {Ricciardi}, {Riller}, {Ristorcelli}, {Rocha}, {Rosset},
  {Rubi{\~n}o-Mart{\'\i}n}, {Rusholme}, {Saar}, {Sandri}, {Santos}, {Saunders},
  {Savini}, {Schaefer}, {Scott}, {Seiffert}, {Shellard}, {Smoot}, {Stanford},
  {Starck}, {Stivoli}, {Stolyarov}, {Stompor}, {Sudiwala}, {Sunyaev}, {Sutton},
  {Sygnet}, {Taburet}, {Tauber}, {Terenzi}, {Toffolatti}, {Tomasi}, {Torre},
  {Tristram}, {Tuovinen}, {Valenziano}, {Vibert}, {Vielva}, {Villa},
  {Vittorio}, {Wade}, {Wandelt}, {Weller}, {White}, {White}, {Yvon}, {Zacchei},
  \& {Zonca}}]{PLANCK2011-ESZ}
{Planck Collaboration}, {Ade}, P.~A.~R., {Aghanim}, N., {et~al.} 2011, \aap,
  536, A8

\bibitem[{{Planck Collaboration} {et~al.}(2016{\natexlab{c}}){Planck
  Collaboration}, {Ade}, {Aghanim}, {Arnaud}, {Ashdown}, {Aumont},
  {Baccigalupi}, {Banday}, {Barreiro}, {Barrena}, {Bartlett}, {Bartolo},
  {Battaner}, {Battye}, {Benabed}, {Beno{\^\i}t}, {Benoit-L{\'e}vy}, {Bernard},
  {Bersanelli}, {Bielewicz}, {Bikmaev}, {B{\"o}hringer}, {Bonaldi}, {Bonavera},
  {Bond}, {Borrill}, {Bouchet}, {Bucher}, {Burenin}, {Burigana}, {Butler},
  {Calabrese}, {Cardoso}, {Carvalho}, {Catalano}, {Challinor}, {Chamballu},
  {Chary}, {Chiang}, {Chon}, {Christensen}, {Clements}, {Colombi}, {Colombo},
  {Combet}, {Comis}, {Couchot}, {Coulais}, {Crill}, {Curto}, {Cuttaia},
  {Dahle}, {Danese}, {Davies}, {Davis}, {de Bernardis}, {de Rosa}, {de Zotti},
  {Delabrouille}, {D{\'e}sert}, {Dickinson}, {Diego}, {Dolag}, {Dole},
  {Donzelli}, {Dor{\'e}}, {Douspis}, {Ducout}, {Dupac}, {Efstathiou},
  {Eisenhardt}, {Elsner}, {En{\ss}lin}, {Eriksen}, {Falgarone}, {Fergusson},
  {Feroz}, {Ferragamo}, {Finelli}, {Forni}, {Frailis}, {Fraisse}, {Franceschi},
  {Frejsel}, {Galeotta}, {Galli}, {Ganga}, {G{\'e}nova-Santos}, {Giard},
  {Giraud-H{\'e}raud}, {Gjerl{\o}w}, {Gonz{\'a}lez-Nuevo}, {G{\'o}rski},
  {Grainge}, {Gratton}, {Gregorio}, {Gruppuso}, {Gudmundsson}, {Hansen},
  {Hanson}, {Harrison}, {Hempel}, {Henrot-Versill{\'e}},
  {Hern{\'a}ndez-Monteagudo}, {Herranz}, {Hildebrandt}, {Hivon}, {Hobson},
  {Holmes}, {Hornstrup}, {Hovest}, {Huffenberger}, {Hurier}, {Jaffe}, {Jaffe},
  {Jin}, {Jones}, {Juvela}, {Keih{\"a}nen}, {Keskitalo}, {Khamitov}, {Kisner},
  {Kneissl}, {Knoche}, {Kunz}, {Kurki-Suonio}, {Lagache}, {Lamarre}, {Lasenby},
  {Lattanzi}, {Lawrence}, {Leonardi}, {Lesgourgues}, {Levrier}, {Liguori},
  {Lilje}, {Linden-V{\o}rnle}, {L{\'o}pez-Caniego}, {Lubin},
  {Mac{\'\i}as-P{\'e}rez}, {Maggio}, {Maino}, {Mak}, {Mandolesi}, {Mangilli},
  {Martin}, {Mart{\'\i}nez-Gonz{\'a}lez}, {Masi}, {Matarrese}, {Mazzotta},
  {McGehee}, {Mei}, {Melchiorri}, {Melin}, {Mendes}, {Mennella}, {Migliaccio},
  {Mitra}, {Miville-Desch{\^e}nes}, {Moneti}, {Montier}, {Morgante},
  {Mortlock}, {Moss}, {Munshi}, {Murphy}, {Naselsky}, {Nastasi}, {Nati},
  {Natoli}, {Netterfield}, {N{\o}rgaard-Nielsen}, {Noviello}, {Novikov},
  {Novikov}, {Olamaie}, {Oxborrow}, {Paci}, {Pagano}, {Pajot}, {Paoletti},
  {Pasian}, {Patanchon}, {Pearson}, {Perdereau}, {Perotto}, {Perrott},
  {Perrotta}, {Pettorino}, {Piacentini}, {Piat}, {Pierpaoli}, {Pietrobon},
  {Plaszczynski}, {Pointecouteau}, {Polenta}, {Pratt}, {Pr{\'e}zeau}, {Prunet},
  {Puget}, {Rachen}, {Reach}, {Rebolo}, {Reinecke}, {Remazeilles}, {Renault},
  {Renzi}, {Ristorcelli}, {Rocha}, {Rosset}, {Rossetti}, {Roudier}, {Rozo},
  {Rubi{\~n}o-Mart{\'\i}n}, {Rumsey}, {Rusholme}, {Rykoff}, {Sandri}, {Santos},
  {Saunders}, {Savelainen}, {Savini}, {Schammel}, {Scott}, {Seiffert},
  {Shellard}, {Shimwell}, {Spencer}, {Stanford}, {Stern}, {Stolyarov},
  {Stompor}, {Streblyanska}, {Sudiwala}, {Sunyaev}, {Sutton}, {Suur-Uski},
  {Sygnet}, {Tauber}, {Terenzi}, {Toffolatti}, {Tomasi}, {Tramonte},
  {Tristram}, {Tucci}, {Tuovinen}, {Umana}, {Valenziano}, {Valiviita}, {Van
  Tent}, {Vielva}, {Villa}, {Wade}, {Wandelt}, {Wehus}, {White}, {Wright},
  {Yvon}, {Zacchei}, \& {Zonca}}]{PLANCK2015-PSZ2}
{Planck Collaboration}, {Ade}, P.~A.~R., {Aghanim}, N., {et~al.}
  2016{\natexlab{c}}, \aap, 594, A27

\bibitem[{{Planck Collaboration} {et~al.}(2016{\natexlab{d}}){Planck
  Collaboration}, {Ade}, {Aghanim}, {Arnaud}, {Ashdown}, {Aumont},
  {Baccigalupi}, {Banday}, {Barreiro}, {Bartlett}, {Bartolo}, {Battaner},
  {Battye}, {Benabed}, {Beno{\^\i}t}, {Benoit-L{\'e}vy}, {Bernard},
  {Bersanelli}, {Bielewicz}, {Bock}, {Bonaldi}, {Bonavera}, {Bond}, {Borrill},
  {Bouchet}, {Boulanger}, {Bucher}, {Burigana}, {Butler}, {Calabrese},
  {Cardoso}, {Catalano}, {Challinor}, {Chamballu}, {Chary}, {Chiang}, {Chluba},
  {Christensen}, {Church}, {Clements}, {Colombi}, {Colombo}, {Combet},
  {Coulais}, {Crill}, {Curto}, {Cuttaia}, {Danese}, {Davies}, {Davis}, {de
  Bernardis}, {de Rosa}, {de Zotti}, {Delabrouille}, {D{\'e}sert}, {Di
  Valentino}, {Dickinson}, {Diego}, {Dolag}, {Dole}, {Donzelli}, {Dor{\'e}},
  {Douspis}, {Ducout}, {Dunkley}, {Dupac}, {Efstathiou}, {Elsner},
  {En{\ss}lin}, {Eriksen}, {Farhang}, {Fergusson}, {Finelli}, {Forni},
  {Frailis}, {Fraisse}, {Franceschi}, {Frejsel}, {Galeotta}, {Galli}, {Ganga},
  {Gauthier}, {Gerbino}, {Ghosh}, {Giard}, {Giraud-H{\'e}raud}, {Giusarma},
  {Gjerl{\o}w}, {Gonz{\'a}lez-Nuevo}, {G{\'o}rski}, {Gratton}, {Gregorio},
  {Gruppuso}, {Gudmundsson}, {Hamann}, {Hansen}, {Hanson}, {Harrison}, {Helou},
  {Henrot-Versill{\'e}}, {Hern{\'a}ndez-Monteagudo}, {Herranz}, {Hildebrandt},
  {Hivon}, {Hobson}, {Holmes}, {Hornstrup}, {Hovest}, {Huang}, {Huffenberger},
  {Hurier}, {Jaffe}, {Jaffe}, {Jones}, {Juvela}, {Keih{\"a}nen}, {Keskitalo},
  {Kisner}, {Kneissl}, {Knoche}, {Knox}, {Kunz}, {Kurki-Suonio}, {Lagache},
  {L{\"a}hteenm{\"a}ki}, {Lamarre}, {Lasenby}, {Lattanzi}, {Lawrence}, {Leahy},
  {Leonardi}, {Lesgourgues}, {Levrier}, {Lewis}, {Liguori}, {Lilje},
  {Linden-V{\o}rnle}, {L{\'o}pez-Caniego}, {Lubin}, {Mac{\'\i}as-P{\'e}rez},
  {Maggio}, {Maino}, {Mandolesi}, {Mangilli}, {Marchini}, {Maris}, {Martin},
  {Martinelli}, {Mart{\'\i}nez-Gonz{\'a}lez}, {Masi}, {Matarrese}, {McGehee},
  {Meinhold}, {Melchiorri}, {Melin}, {Mendes}, {Mennella}, {Migliaccio},
  {Millea}, {Mitra}, {Miville-Desch{\^e}nes}, {Moneti}, {Montier}, {Morgante},
  {Mortlock}, {Moss}, {Munshi}, {Murphy}, {Naselsky}, {Nati}, {Natoli},
  {Netterfield}, {N{\o}rgaard-Nielsen}, {Noviello}, {Novikov}, {Novikov},
  {Oxborrow}, {Paci}, {Pagano}, {Pajot}, {Paladini}, {Paoletti}, {Partridge},
  {Pasian}, {Patanchon}, {Pearson}, {Perdereau}, {Perotto}, {Perrotta},
  {Pettorino}, {Piacentini}, {Piat}, {Pierpaoli}, {Pietrobon}, {Plaszczynski},
  {Pointecouteau}, {Polenta}, {Popa}, {Pratt}, {Pr{\'e}zeau}, {Prunet},
  {Puget}, {Rachen}, {Reach}, {Rebolo}, {Reinecke}, {Remazeilles}, {Renault},
  {Renzi}, {Ristorcelli}, {Rocha}, {Rosset}, {Rossetti}, {Roudier},
  {Rouill{\'e} d'Orfeuil}, {Rowan-Robinson}, {Rubi{\~n}o-Mart{\'\i}n},
  {Rusholme}, {Said}, {Salvatelli}, {Salvati}, {Sandri}, {Santos},
  {Savelainen}, {Savini}, {Scott}, {Seiffert}, {Serra}, {Shellard}, {Spencer},
  {Spinelli}, {Stolyarov}, {Stompor}, {Sudiwala}, {Sunyaev}, {Sutton},
  {Suur-Uski}, {Sygnet}, {Tauber}, {Terenzi}, {Toffolatti}, {Tomasi},
  {Tristram}, {Trombetti}, {Tucci}, {Tuovinen}, {T{\"u}rler}, {Umana},
  {Valenziano}, {Valiviita}, {Van Tent}, {Vielva}, {Villa}, {Wade}, {Wandelt},
  {Wehus}, {White}, {White}, {Wilkinson}, {Yvon}, {Zacchei}, \&
  {Zonca}}]{PLANCK2016-Cosmology}
{Planck Collaboration}, {Ade}, P.~A.~R., {Aghanim}, N., {et~al.}
  2016{\natexlab{d}}, \aap, 594, A13

\bibitem[{{Planck Collaboration} {et~al.}(2016{\natexlab{e}}){Planck
  Collaboration}, {Ade}, {Aghanim}, {Arnaud}, {Ashdown}, {Aumont},
  {Baccigalupi}, {Banday}, {Barreiro}, {Bartlett}, {Bartolo}, {Battaner},
  {Battye}, {Benabed}, {Beno{\^\i}t}, {Benoit-L{\'e}vy}, {Bernard},
  {Bersanelli}, {Bielewicz}, {Bock}, {Bonaldi}, {Bonavera}, {Bond}, {Borrill},
  {Bouchet}, {Bucher}, {Burigana}, {Butler}, {Calabrese}, {Cardoso},
  {Catalano}, {Challinor}, {Chamballu}, {Chary}, {Chiang}, {Christensen},
  {Church}, {Clements}, {Colombi}, {Colombo}, {Combet}, {Comis}, {Couchot},
  {Coulais}, {Crill}, {Curto}, {Cuttaia}, {Danese}, {Davies}, {Davis}, {de
  Bernardis}, {de Rosa}, {de Zotti}, {Delabrouille}, {D{\'e}sert}, {Diego},
  {Dolag}, {Dole}, {Donzelli}, {Dor{\'e}}, {Douspis}, {Ducout}, {Dupac},
  {Efstathiou}, {Elsner}, {En{\ss}lin}, {Eriksen}, {Falgarone}, {Fergusson},
  {Finelli}, {Forni}, {Frailis}, {Fraisse}, {Franceschi}, {Frejsel},
  {Galeotta}, {Galli}, {Ganga}, {Giard}, {Giraud-H{\'e}raud}, {Gjerl{\o}w},
  {Gonz{\'a}lez-Nuevo}, {G{\'o}rski}, {Gratton}, {Gregorio}, {Gruppuso},
  {Gudmundsson}, {Hansen}, {Hanson}, {Harrison}, {Henrot-Versill{\'e}},
  {Hern{\'a}ndez-Monteagudo}, {Herranz}, {Hildebrandt}, {Hivon}, {Hobson},
  {Holmes}, {Hornstrup}, {Hovest}, {Huffenberger}, {Hurier}, {Jaffe}, {Jaffe},
  {Jones}, {Juvela}, {Keih{\"a}nen}, {Keskitalo}, {Kisner}, {Kneissl},
  {Knoche}, {Kunz}, {Kurki-Suonio}, {Lagache}, {L{\"a}hteenm{\"a}ki},
  {Lamarre}, {Lasenby}, {Lattanzi}, {Lawrence}, {Leonardi}, {Lesgourgues},
  {Levrier}, {Liguori}, {Lilje}, {Linden-V{\o}rnle}, {L{\'o}pez-Caniego},
  {Lubin}, {Mac{\'\i}as-P{\'e}rez}, {Maggio}, {Maino}, {Mandolesi}, {Mangilli},
  {Maris}, {Martin}, {Mart{\'\i}nez-Gonz{\'a}lez}, {Masi}, {Matarrese},
  {McGehee}, {Meinhold}, {Melchiorri}, {Melin}, {Mendes}, {Mennella},
  {Migliaccio}, {Mitra}, {Miville-Desch{\^e}nes}, {Moneti}, {Montier},
  {Morgante}, {Mortlock}, {Moss}, {Munshi}, {Murphy}, {Naselsky}, {Nati},
  {Natoli}, {Netterfield}, {N{\o}rgaard-Nielsen}, {Noviello}, {Novikov},
  {Novikov}, {Oxborrow}, {Paci}, {Pagano}, {Pajot}, {Paoletti}, {Partridge},
  {Pasian}, {Patanchon}, {Pearson}, {Perdereau}, {Perotto}, {Perrotta},
  {Pettorino}, {Piacentini}, {Piat}, {Pierpaoli}, {Pietrobon}, {Plaszczynski},
  {Pointecouteau}, {Polenta}, {Popa}, {Pratt}, {Pr{\'e}zeau}, {Prunet},
  {Puget}, {Rachen}, {Rebolo}, {Reinecke}, {Remazeilles}, {Renault}, {Renzi},
  {Ristorcelli}, {Rocha}, {Roman}, {Rosset}, {Rossetti}, {Roudier},
  {Rubi{\~n}o-Mart{\'\i}n}, {Rusholme}, {Sandri}, {Santos}, {Savelainen},
  {Savini}, {Scott}, {Seiffert}, {Shellard}, {Spencer}, {Stolyarov}, {Stompor},
  {Sudiwala}, {Sunyaev}, {Sutton}, {Suur-Uski}, {Sygnet}, {Tauber}, {Terenzi},
  {Toffolatti}, {Tomasi}, {Tristram}, {Tucci}, {Tuovinen}, {T{\"u}rler},
  {Umana}, {Valenziano}, {Valiviita}, {Van Tent}, {Vielva}, {Villa}, {Wade},
  {Wandelt}, {Wehus}, {Weller}, {White}, {Yvon}, {Zacchei}, \&
  {Zonca}}]{PLANCK2016-PSZ2_cosmology}
{Planck Collaboration}, {Ade}, P.~A.~R., {Aghanim}, N., {et~al.}
  2016{\natexlab{e}}, \aap, 594, A24

\bibitem[{{Planck Collaboration} {et~al.}(2016{\natexlab{f}}){Planck
  Collaboration}, {Aghanim}, {Arnaud}, {Ashdown}, {Aumont}, {Baccigalupi},
  {Banday}, {Barreiro}, {Bartlett}, {Bartolo}, {Battaner}, {Battye}, {Benabed},
  {Beno{\^\i}t}, {Benoit-L{\'e}vy}, {Bernard}, {Bersanelli}, {Bielewicz},
  {Bock}, {Bonaldi}, {Bonavera}, {Bond}, {Borrill}, {Bouchet}, {Burigana},
  {Butler}, {Calabrese}, {Cardoso}, {Catalano}, {Challinor}, {Chiang},
  {Christensen}, {Churazov}, {Clements}, {Colombo}, {Combet}, {Comis},
  {Coulais}, {Crill}, {Curto}, {Cuttaia}, {Danese}, {Davies}, {Davis}, {de
  Bernardis}, {de Rosa}, {de Zotti}, {Delabrouille}, {D{\'e}sert}, {Dickinson},
  {Diego}, {Dolag}, {Dole}, {Donzelli}, {Dor{\'e}}, {Douspis}, {Ducout},
  {Dupac}, {Efstathiou}, {Elsner}, {En{\ss}lin}, {Eriksen}, {Fergusson},
  {Finelli}, {Forni}, {Frailis}, {Fraisse}, {Franceschi}, {Frejsel},
  {Galeotta}, {Galli}, {Ganga}, {G{\'e}nova-Santos}, {Giard},
  {Gonz{\'a}lez-Nuevo}, {G{\'o}rski}, {Gregorio}, {Gruppuso}, {Gudmundsson},
  {Hansen}, {Harrison}, {Henrot-Versill{\'e}}, {Hern{\'a}ndez-Monteagudo},
  {Herranz}, {Hildebrandt}, {Hivon}, {Holmes}, {Hornstrup}, {Huffenberger},
  {Hurier}, {Jaffe}, {Jones}, {Juvela}, {Keih{\"a}nen}, {Keskitalo}, {Kneissl},
  {Knoche}, {Kunz}, {Kurki-Suonio}, {Lacasa}, {Lagache}, {L{\"a}hteenm{\"a}ki},
  {Lamarre}, {Lasenby}, {Lattanzi}, {Leonardi}, {Lesgourgues}, {Levrier},
  {Liguori}, {Lilje}, {Linden-V{\o}rnle}, {L{\'o}pez-Caniego},
  {Mac{\'\i}as-P{\'e}rez}, {Maffei}, {Maggio}, {Maino}, {Mandolesi},
  {Mangilli}, {Maris}, {Martin}, {Mart{\'\i}nez-Gonz{\'a}lez}, {Masi},
  {Matarrese}, {Melchiorri}, {Melin}, {Migliaccio}, {Miville-Desch{\^e}nes},
  {Moneti}, {Montier}, {Morgante}, {Mortlock}, {Munshi}, {Murphy}, {Naselsky},
  {Nati}, {Natoli}, {Noviello}, {Novikov}, {Novikov}, {Paci}, {Pagano},
  {Pajot}, {Paoletti}, {Pasian}, {Patanchon}, {Perdereau}, {Perotto},
  {Pettorino}, {Piacentini}, {Piat}, {Pierpaoli}, {Pietrobon}, {Plaszczynski},
  {Pointecouteau}, {Polenta}, {Ponthieu}, {Pratt}, {Prunet}, {Puget}, {Rachen},
  {Reinecke}, {Remazeilles}, {Renault}, {Renzi}, {Ristorcelli}, {Rocha},
  {Rossetti}, {Roudier}, {Rubi{\~n}o-Mart{\'\i}n}, {Rusholme}, {Sandri},
  {Santos}, {Sauv{\'e}}, {Savelainen}, {Savini}, {Scott}, {Spencer},
  {Stolyarov}, {Stompor}, {Sunyaev}, {Sutton}, {Suur-Uski}, {Sygnet}, {Tauber},
  {Terenzi}, {Toffolatti}, {Tomasi}, {Tramonte}, {Tristram}, {Tucci},
  {Tuovinen}, {Valenziano}, {Valiviita}, {Van Tent}, {Vielva}, {Villa}, {Wade},
  {Wandelt}, {Wehus}, {Yvon}, {Zacchei}, \& {Zonca}}]{PLANCK2016-ymap}
{Planck Collaboration}, {Aghanim}, N., {Arnaud}, M., {et~al.}
  2016{\natexlab{f}}, \aap, 594, A22

\bibitem[{{Pointecouteau} {et~al.}(2021){Pointecouteau}, {Santiago-Bautista},
  {Douspis}, {Aghanim}, {Crichton}, {Diego}, {Hurier}, {Macias-Perez},
  {Marriage}, {Remazeilles}, {Caretta}, \&
  {Bravo-Alfaro}}]{Pointecouteau2021-profile-PACT}
{Pointecouteau}, E., {Santiago-Bautista}, I., {Douspis}, M., {et~al.} 2021,
  \aap, 651, A73

\bibitem[{{Rozo} {et~al.}(2010){Rozo}, {Wechsler}, {Rykoff}, {Annis}, {Becker},
  {Evrard}, {Frieman}, {Hansen}, {Hao}, {Johnston}, {Koester}, {McKay},
  {Sheldon}, \& {Weinberg}}]{Rozo2010-SDSS_cluster_cosmology}
{Rozo}, E., {Wechsler}, R.~H., {Rykoff}, E.~S., {et~al.} 2010, \apj, 708, 645

\bibitem[{{Schneider} \& {Bartelmann}(1997)}]{schneider1997}
{Schneider}, P. \& {Bartelmann}, M. 1997, \mnras, 286, 696

\bibitem[{{Springel}(2010)}]{springel2010}
{Springel}, V. 2010, \mnras, 401, 791

\bibitem[{{Springel} {et~al.}(2018){Springel}, {Pakmor}, {Pillepich},
  {Weinberger}, {Nelson}, {Hernquist}, {Vogelsberger}, {Genel}, {Torrey},
  {Marinacci}, \& {Naiman}}]{TNG(c)}
{Springel}, V., {Pakmor}, R., {Pillepich}, A., {et~al.} 2018, \mnras, 475, 676

\bibitem[{{Steigman}(2008)}]{Steigman2008-BBN}
{Steigman}, G. 2008, arXiv e-prints, arXiv:0807.3004

\bibitem[{{Sunyaev} \& {Zeldovich}(1980)}]{Sunyaev-Zeldovich1980}
{Sunyaev}, R.~A. \& {Zeldovich}, I.~B. 1980, \araa, 18, 537

\bibitem[{{Sunyaev} \& {Zeldovich}(1970)}]{Sunyaev-Zeldovich1970}
{Sunyaev}, R.~A. \& {Zeldovich}, Y.~B. 1970, \apss, 7, 3

\bibitem[{{Sunyaev} \& {Zeldovich}(1972)}]{Sunyaev-Zeldovich1972}
{Sunyaev}, R.~A. \& {Zeldovich}, Y.~B. 1972, Comments on Astrophysics and Space
  Physics, 4, 173

\bibitem[{{Tramonte} {et~al.}(2023){Tramonte}, {Ma}, {Yan}, {Maturi},
  {Castignani}, {Sereno}, {Bardelli}, {Giocoli}, {Marulli}, {Moscardini},
  {Puddu}, {Radovich}, {Van Waerbeke}, \&
  {Wright}}]{Tramonte2023-profile-stacked}
{Tramonte}, D., {Ma}, Y.-Z., {Yan}, Z., {et~al.} 2023, arXiv e-prints,
  arXiv:2302.06266

\bibitem[{{Vall{\'e}s-P{\'e}rez} {et~al.}(2020){Vall{\'e}s-P{\'e}rez},
  {Planelles}, \& {Quilis}}]{Vallesperez2020}
{Vall{\'e}s-P{\'e}rez}, D., {Planelles}, S., \& {Quilis}, V. 2020, \mnras, 499,
  2303

\bibitem[{{Zubeldia} {et~al.}(2022){Zubeldia}, {Rotti}, {Chluba}, \&
  {Battye}}]{Zubeldia2022a-iMMF}
{Zubeldia}, {\'I}., {Rotti}, A., {Chluba}, J., \& {Battye}, R. 2022, arXiv
  e-prints, arXiv:2204.13780

\end{thebibliography}

\end{document}